\documentstyle[12pt,epsf]{article}

\begin{document}
%%%%%%%%%%%%%%%%%%%%%%%%%%%%%%%%%%%%%%%%%%%%%%%%%%%%%%%%%%%%%%%%%%%%%%%%%%
\newcommand{\bda}{\begin{\displaymath}\begin{array}{rl}}
\newcommand{\eda}{\end{array}\end{displaymath}}
\newcommand{\be}{\begin{equation}}
\newcommand{\ee}{\end{equation}}
\newcommand{\bdm}{\begin{displaymath}}
\newcommand{\edm}{\end{displaymath}}
\newcommand{\bea}{\begin{eqnarray}}
\newcommand{\eea}{\end{eqnarray}}
\newcommand{\bfm}{\boldmath}
\newcommand{\bfpi}{{\mbox{\boldmath{$\pi$}}}}
\newcommand{\bfphi}{{\mbox{\boldmath{$\phi$}}}}
\newcommand{\bftau}{{\mbox{\boldmath{$\tau$}}}}
\newcommand{\no}{\nonumber \\}
\newcommand{\fs}{\; \; .}
\newcommand{\co}{\; \; ,}
\newcommand{\per}{\;\;.}
\newcommand{\rar}{\rightarrow}
\newcommand{\nn}{\nonumber \\}
\newcommand{\mtiny}[1]{{\mbox{\tiny #1}}}
\newcommand{\al}{&\!\!\!\!}
\newcommand{\eff}{e\hspace{-0.1em}f\hspace{-0.18em}f}
\newcommand{\free}{\!f\hspace{-0.05em}r\hspace{-0.05em}e\hspace{-0.02em}e}
\newcommand{\ind}{\scriptscriptstyle}
\newcommand{\gra}{{\scriptscriptstyle\gamma}}
\newcommand{\QCD}{\mbox{\scriptsize Q\hspace{-0.1em}CD}}
\newcommand{\indR}{\mbox{\tiny R}}
\newcommand{\indL}{\mbox{\tiny L}}
\newcommand{\indV}{\mbox{\tiny V}}
\newcommand{\lvac}{\langle 0|\,}
\newcommand{\rvac}{\,|0\rangle}
\newcommand{\lav}{\langle\hspace{-0.2em}\langle}
\newcommand{\rav}{\rangle\hspace{-0.2em}\rangle}
\newcommand{\wave}{\raisebox{0.22em}{\fbox{\rule[0.15em]{0em}{0em}\,}}\,}
\newcommand{\mvec}{\vec{\rule{0em}{0.6em}m}}
\newcommand{\sm}{$\sigma$-model }
\newcommand{\permille}{\%\raisebox{-0.05em}{{\scriptsize 0}}\,}
\newcommand{\LWZ}{{\cal L}_{\scriptscriptstyle W\hspace{-0.1em}Z}}
\newcounter{figuren}
\setcounter{figuren}{1}
\renewcommand{\thefigure}{\thefiguren}
%%%%%%%%%%%%%%%%%%%%%%%%%%%%%%%%%%%%%%%%%%%%%%%%%%%%%%%%%%%%%%%%%%%%%%%%%%%%
\begin{titlepage}
\begin{flushright}BUTP-94/13\end{flushright}
\rule{0em}{2em}\vspace{4em}
\begin{center}
{\LARGE {\bf Principles of \\\rule{0em}{1em}Chiral Perturbation Theory}}\\
\vspace{2em}
H.Leutwyler\\Institut f\"{u}r theoretische Physik der Universit\"{a}t Bern\\
Sidlerstr. 5, CH-3012 Bern, Switzerland\\

{\scriptsize HLEUTWYLER@ITP.UNIBE.CH}

\vspace{5em}
May 1994\\
\vspace{7em}

Lectures given at the Workshop "Hadrons 1994", Gramado, RS, Brasil

\vspace{8em}
\rule{30em}{.02em}\\
{\footnotesize Work
supported in part by Schweizerischer Nationalfonds}
\end{center}
\end{titlepage}
%%%%%%%%%%%%%%%%%%%%%%%%%%%%%%%%%%%%%%%%%%%%%%%%%%%%%%%%%%%%%%%%%%%
\renewcommand{\baselinestretch}{0.5}
{\footnotesize
\tableofcontents}
\normalsize
\renewcommand{\baselinestretch}{1}
%%%%%%%%%%%%%%%%%%%%%%%%%%%%%%%%%%%%%%%%%%%%%%%%%%%%%%%%%%%%%%%%%%%%
\section{Introduction}
\label{intro}
An elementary discussion of the main concepts used in chiral perturbation
theory is given in textbooks
\cite{Georgi,Donoghue Golowich Holstein} and a more detailed picture of the
applications may be obtained from the reviews listed in
\cite{Schladming,Meissner,Ecker Cargese}. For an
overview of ongoing work, I refer
to \cite{Budapest,Dallas,Dafne}. Concerning
the foundations of the method, however, the literature
is comparatively scarce. So, I will concentrate on the basic concepts and
explain {\it why} the method works.

Chiral perturbation theory ($\chi$PT) is an {\it effective field theory}. The
main application is QCD, where the method leads to a rather detailed
and quantitative understanding of the low energy structure. Despite its name,
$\chi$PT is a nonperturbative method, because
it does not rely on an expansion in powers of the QCD coupling constant. The
method invokes a different expansion: it makes use of the fact that,
at low energies, the behaviour of scattering amplitudes or current
matrix
elements can be described in terms of a {\it Taylor series expansion} in powers
of the momenta.
The electromagnetic form factor of the pion, e.g., may be
exanded in powers of the momentum transfer $t$.
In this case, the first two Taylor coefficients are related to the total charge
of the particle and to the mean square radius of the charge distribution,
respectively,
\be \label{taylor}
f_{\pi^+}(t) = 1 + \mbox{$\frac{1}{6}$} \langle r^2\rangle_{\pi^+}\, t +
O(t^2)\fs \end{equation}
Scattering lengths and effective ranges are analogous low energy
constants occurring in the Taylor series expansion of scattering amplitudes.

For the straightforward expansion in powers of the momenta to hold, it is
essential that the theory does not contain massless particles. The exchange of
photons, e.g., gives rise to Coulomb scattering, described by an amplitude of
the form $e^2/(p'-p)^2$, which does not admit a Taylor series expansion. Now,
QCD does not contain massless particles, but it does contain very light ones:
pions. The occurrence of light particles gives rise to singularities in the low
energy domain, which limit the range of validity of the Taylor series
representation. The form factor $f_{\pi^+}(t)$, e.g., contains a branch point
singularity
at $t=4 M_\pi^2$, such that the formula (\ref{taylor}) provides an adequate
representation only for $t\ll 4 M_\pi^2$. To extend this representation to
larger momenta, one needs to account for the singularities generated by the
pions. This can be done, because the reason why $M_\pi$ is so small is
understood: the pions are the Goldstone bosons of a hidden, approximate
symmetry \cite{Nambu}.

The main consequences of this
symmetry were derived in the sixties, from a direct analysis
of the Ward identities, using current algebra and pion pole dominance.
$\chi$PT addresses the same problem
in a more systematic manner and is considerably more efficient
\cite{Weinberg1979,GLAnnals,GLNP,found}.
The corresponding series expansion amounts to
a modified Taylor
series, which explicitly accounts for the singularities generated by the
Goldstone bosons.
It provides a solid mathematical basis for what used to be called the "PCAC
hypothesis".
%%%%%%%%%%%%%%%%%%%%%%%%%%%%%%%%%%%%%%%%%%%%%%%%%%%%%%%%%%%%%%%%%%%%%%%%
\section{Goldstone theorem}
\label{gt}
To start, let me briefly review the notion of a {\it
spontaneously broken
symmetry}.
Consider any field theory model, for which the Hamiltonian is
invariant under
some Lie group G. Denote the generators of this group by $Q_i$, such that
\bdm
[Q_i, \mbox{\bf H}]=  0\fs
\edm
The symmetry is called spontaneously broken if the ground state of the theory
is not invariant under G. Suppose, therefore, that, for some of the
generators
\bdm
Q_i \rvac \neq 0\fs\edm
This immediately implies that the vacuum is not the only state of zero energy:
since {\bf H} commutes with $Q_i$, the vector $Q_i \rvac$ describes a state
with
the same energy as the vacuum. In a relativistically invariant theory, this can
only happen if the spectrum of physical states contains massless particles,
Goldstone bosons.

The subset formed by those generators,
which do leave the ground state invariant, is a subalgebra: if $Q_i$ and $Q_k$
annihilate the vacuum, then this is also true of the commutator $[Q_i, Q_k]$.
These operators therefore generate a subgroup $\mbox{H} \subset \mbox{G}$.
Spontaneous symmetry breakdown thus involves two groups --- the
symmetry group G of the Hamiltonian and the symmetry group H of the vacuum.
Denote the number of parameters required to label the elements of G by $n_G$
such that there are $n_G$ generators and suppose that $n_H < n_G$ is the number
of parameters occurring in H. The $n_G - n_H$ generators which belong to the
quotient G/H of the two groups do not annihilate the ground state. The
corresponding vectors $Q_i \rvac$ are linearly independent, because,
otherwise, a suitable linear combination of these generators would leave the
vacuum invariant and hence belong to H. Accordingly, spontaneous breakdown of
the group G to the subgroup H requires the occurrence of $n_G -n_H$
independent states of zero energy: the spectrum of the theory must contain
$n_G-n_H$ different flavours of Goldstone bosons. The Goldstone theorem
amounts to a mathematically precise formulation of this statement
\cite{Goldstone,Coleman}.
%%%%%%%%%%%%%%%%%%%%%%%%%%%%%%%%%%%%%%%%%%%%%%%%%%%%%%%%%%%%%%%%%%%%%%%%
\section{QCD}
\label{qcd}
In the case of QCD, the relevant spontaneously broken symmetry is
an approximate one,
related to the occurrence of several quark flavours. I denote the
quark field by $q(x)$,
\bdm
\label{qcd1}
q(x) =\pmatrix{u(x) \cr
d(x)  \cr
s(x) \cr
\vdots \cr
}
\edm
The Lagrangian is of the form
\bdm
\label{qcd2}
{\cal L}_{\QCD} = - \frac{1}{2g^2} \mbox{tr}_cG_{\mu\nu} G^{\mu\nu} + \bar{q} i
\gamma^\mu D_\mu q - \bar{q}\, m\,q \co
\edm
where $G_{\mu \nu}$ is the field strength of the gluon field,
\bdm
\label{qcd3}
G_{\mu \nu} = \partial_\mu G_\nu - \partial_\nu G_\mu
-i [G_\mu ,G_\nu ] \co
\edm
$D_\mu $ denotes the covariant derivative,
\bdm
\label{qcd4}
D_\mu q(x) = \partial _\mu q(x) - i G_\mu (x) q(x)
\edm
and $m$ is the quark mass matrix,
\bdm
\label{qcd5}
m=\left(\;\raisebox{1.5em}{$m_u$}\;\raisebox{0.5em}{$m_d$}\;
\raisebox{-0.5em}{$m_s$}\;\raisebox{-1.8em}{$\ddots\;$}\, \right)
\edm
Note that the coupling constant $g$ is absorbed in the gluon
field: in the notation used here, the covariant derivative involves $G_\mu$
rather
than $gG_\mu$. In the Lagrangian, the coupling constant then only occurs in
front of the term proportional to the square of the field strength. Also,
colour indices are suppressed --- the symbol tr$_c$ denotes the trace
of a colour matrix.

The basic parameters of QCD are the dimensionless bare coupling constant $g$
and the bare quark mass matrix $m$. Both of these must be
tuned to the magnitude of the cutoff $\mu$ for the limit
$\mu\rightarrow \infty$ to make sense: $g=g(\mu),\,m=m(\mu)$.
In order for
a change in $\mu$ not to modify physical quantities, such
as bound state masses, the bare coupling constant must be shifted by an amount
determined by the $\beta
$-function,
\bdm\label{qcd5a} \mu \frac{dg}{d\mu } = \beta (g)\fs\edm
At small coupling, $\beta$ is negative, i.e., the theory is asymptotically
free.
In the minimal subtraction scheme, the $\beta$-function is independent of the quark
masses. The leading term in the perturbative expansion is of order $g^3$,
\bdm
\label{qcd5b}
\beta (g) =
-\beta_0\frac{g^3}{(4\pi )^2}
+O(g^5)\co\hspace{3em}
\beta_0=
11 - \mbox{$\frac{2}{3}$}\mbox{N}_f \co\edm
where $\mbox{N}_f$ is the number of quark flavours. With this expression for
the rhs, the above differential
equation is readily integrated. The result for $g(\mu)^2$ is
inversely proportional to the logarithm of the cutoff,
\bdm\label{qcd5e}\frac{g(\mu)^2}{(4\pi )^2} = \frac{1}{\beta_0\,
\ln (\mu^2/\Lambda_{\QCD}^{\hspace{1.6em}2})}\co
\edm
where $\Lambda_{\QCD}$ is the constant of integration. By definition,
this quantity is independent
of the cutoff and thus represents a significant
parameter, referred to as the renormalization
group invariant scale of QCD. The theory is not
characterized by the dimensionless coupling constant $g$ occurring in the
Lagrangian
--- the value of this parameter depends on the cutoff --- but by the mass scale
$\Lambda_{\QCD}$ ("dimensional transmutation").

The tuning of the quark mass matrix is determined by
the $\gamma$-function,
\bdm\label{qcd5y}
\mu \frac{dm}{d\mu } =- \gamma (g)\,m\hspace{1em},\hspace{1em}
\gamma(g)=\frac{g^2}{2\pi^2}+O(g^4)\edm
For large values of $\mu$, where it is justified to only retain the leading
terms of the perturbative expansion, the solution is of the form
\bdm\label{qcd5f}
m(\mu) = \{\ln \frac{\mu}{\Lambda_{\QCD}}\}^{-\frac{4}{\beta_0}}
\;\bar{m}
\fs\edm
The constant of integration $\,\bar{m}\,$ occurring here is the
renormalization group invariant quark
mass matrix. When the cutoff is sent to infinity, the bare constants $g$ and
$m$ tend to zero, roughly like $g\sim (\ln \mu )^{-1},\;m\sim (\ln
\mu )^{-\frac{1}{2}}$, while the observables of the theory
(hadron masses, decay constants, scattering amplitudes etc.) approach finite
limits,
determined by the renormalization group invariant quantities
$\Lambda_{\QCD},\, \bar{m}_u,\,\bar{m}_d,
\ldots $

The differential equations for the running coupling constant and running
quark masses determine the magnitude of these quantities
also for values of $\mu$ which are not large compared to $\Lambda_{\QCD}$,
but
the first one or two terms in the perturbative expansion of the functions
$\beta$ and $\gamma$ do then not provide an adequate representation.
One refers
to the functions $g(\mu)$ and $m(\mu)$, defined by these equations, as the
running coupling constant and running quark mass, respectively and calls the
parameter $\mu$
the running scale. The scale may be given any value --- the observables are
independent thereof. If $\mu$ is large, the running coupling constant
becomes small, such that the scale dependence is controlled by the perturbation
theory formulae given above.

The vector
and axial currents are not renormalized. Despite the fact that the dimension
of the
singlet axial current is anomalous, it does not require wave
function renormalization.
The scalar and pseudoscalar quark
densities, on the other hand, need to be renormalized, in order for their
Green functions
to approach finite limits when the cutoff is removed. The relevant
$Z$-factor is the inverse of the one occurring in the
quark mass matrix --- the products $m\,\bar{q_1}{q_2}$ and
$m\,\bar{q_1}\gamma_5q_2$ are renormalization group invariant.
In the following, I will
throughout be working with the running
quark masses and densities, without
explicitly indicating that
these quantitites depend on $\mu$. Equally well, we
may
use the renormalization group invariant quark masses $\bar{m}$,
provided we use the same convention also when normalizing the scalar
and pseudoscalar operators.
%%%%%%%%%%%%%%%%%%%%%%%%%%%%%%%%%%%%%%%%%%%%%%%%%%%%%%%%%%%%%%%%%%%%%%
\section{Massless quarks}\label{qcdmq}
As far as the strong interactions are concerned, the different quarks
$u,\,d,\,\ldots\,$ have identical properties, except for their mass.
From a theoretical point of view, the quark masses
represent free parameters of the QCD
Lagrangian. The theory makes sense for any value of $m_u, m_d, m_s, \ldots\,$
We now first consider the fictitious world where all of the quarks
are taken massless. This world
is a theoretician's paradise: a theory
without adjustable dimensionless parameters whatsoever
(although the
Lagrangian
does contain one dimensionless coupling constant $g$, the value of this
constant is without significance, as it merely determines the running
scale in units of the renormalization group invariant scale $\Lambda_{\QCD}$).

If the quarks are massless,
the Lagrangian does not contain any terms which connect the
right- and left-handed components of the quark fields,
\bdm q_{\indR} = \mbox{$ \frac{1}{2}$} (1+ \gamma_5)q\;\;,\;\;
q_{\indR} =\mbox{$\frac{1}{2}$} (1- \gamma_5)q\fs\edm
The Lagrangian of massless QCD,
therefore, remains invariant under "chiral" rotations, i.e., under independent
transformations of the right- and left-handed quark fields,
\bdm
\label{qcd6}
q_{\indR} \rightarrow
V_{\indR} q_{\indR}\;\;,\;\;
q_{\indL}
\rightarrow V_{\indL} q_{\indL}
\hspace{4em} V_{\indR},\,V_{\indL}\in
\mbox{U(N}_f)\fs \edm
The Noether
currents associated with this symmetry of the Lagrangian are given by
\bdm\begin{array}{lllllll}
\label{qcd7}
V_a^\mu&=&\bar{q} \gamma^\mu \mbox{$\frac{1}{2}$}\lambda_a q\;,&\;
A_a^\mu &=& \bar{q} \gamma^\mu \gamma_5 \mbox{$\frac{1}{2}$} \lambda_a q\;,
&\;\; a=1,\ldots\,,\mbox{N}_f^2\!-\!1\\
V_0^\mu&=&\bar{q} \gamma^\mu q\;,&
A_0^\mu &=& \bar{q} \gamma^\mu \gamma_5 q\;, &\end{array}\edm
where the Gell-Mann matrices $\lambda_1,\,\lambda_2,\,\ldots$ form
a complete set of traceless, hermitean $\mbox{N}_f\!\times\!\mbox{N}_f$
matrices.

One of these currents, however, is anomalous: despite the symmetry of the
Lagrangian, the singlet axial current $A_0^\mu$ fails to be
conserved (see section \ref{anom}),
\bdm \partial_\mu A^\mu_0=
\frac{\mbox{N}_f}{8\pi ^2} \mbox{tr}_c G_{\mu \nu} \tilde{G}^{\mu \nu }\fs\edm
The actual symmetry group of mass\-less QCD is gene\-rated by the char\-ges of
the conserved currents $V_a^\mu, V_0^\mu$ and $A_a^\mu$. It consists of
those pairs of elements $V_{\indR}, V_{\indL}
\in \mbox{U}(\mbox{N}_f)$ which obey the
constraint
$\det (V_{\indR} V_{\indL}^{-1}) = 1$,
i.e.,
\bdm
\label{qcd10}
\mbox{G}_0 = \mbox{SU}(\mbox{N}_f)_{\indR} \times
\mbox{SU}(\mbox{N}_f)_{\indL}
\times \mbox{U}(1)_{\indV} \fs
\edm
%%%%%%%%%%%%%%%%%%%%%%%%%%%%%%%%%%%%%%%%%%%%%%%%%%%%%%%%%%%%%%%%%%%%%%%%%%
\section{Spontaneous breakdown of chiral symmetry}
\label{qcdsb}
For QCD to describe the strong interactions observed in nature, it is crucial
that chiral symmetry is spontaneously broken, the ground state
being invariant only under the subgroup
generated by the charges of the vector currents. There are theoretical
arguments indicating that chromodynamics indeed leads to the formation of a
quark condensate, which is invariant under the subgroup generated by
the vector charges, but correlates the
right- and lefthanded fields and thus breaks chiral invariance \cite{Witten}.
The available lattice results also support the hypothesis. Taking the
generally accepted picture for granted,
the vacuum is invariant only under
the subgroup
\bdm
\label{qcd11}
\mbox{H}_0 = \mbox{SU}(\mbox{N}_f)_{\indV}\times
\mbox{U}(1)_{\indV} \fs \edm

The spontaneous symmetry breakdown gives rise to
$\mbox{N}_f^2\!-\!1$ Goldstone bosons,
where $\mbox{N}_f$ is the number of quark flavours. So, the spectrum of QCD
with $\mbox{N}_f>1$ massless quarks must contain
$\mbox{N}_f^2\!-\!1$ massless physical states. Their quantum numbers coincide
with those of the states obtained by applying the axial
charge operators
to the vacuum: $J^P= 0^-$.

The factor $\mbox{U(1)}_{\indV}$ which occurs in both
$\mbox{G}_0$ and $\mbox{H}_0$ is generated by
the charge belonging to the singlet vector current $V_0^\mu$. This charge
counts the number of quarks minus the number of antiquarks:
$3V_0^\mu$ is the current belonging to baryon number.

If the vacuum was symmetric with respect to $\mbox{G}_0$, only those
operators, which are invariant under this group, could pick up a nonzero
vacuum expectation value. For a spontaneously broken symmetry, however,
this does not hold. One refers to the vacuum expectation values of
operators which transform in a nontrivial manner under the
symmetry group as {\it order parameters}.
Since the vacuum
is invariant both under Lorentz transformations and under space reflections,
only scalar operators can develop nonzero vacuum expectation values.
Furthermore, the symmetry of the vacuum insures that only H-invariant
operators can give rise to order parameters. In QCD,
the scalar operator of lowest dimension which qualifies is $\bar{q}q$.
The corresponding order parameter,
$\lvac \bar{q} q \rvac$,
is referred to as the {\it quark condensate}. The operator
$\bar{q}\lambda_aq$ cannot develop a vacuum expectation value, because
it is not invariant under H. In the massless theory, the different flavours
thus all pick up the same expectation value,
\bdm \lvac\bar{u}u\rvac=\lvac\bar{d}d\rvac=
\lvac\bar{s}s\rvac=\ldots\edm

The right-handed quark field $q_{\mbox{\scriptsize R}}$
transforms
according to the fundamental representation $\mbox{\underline{N}}_f$ of
$\mbox{SU}(\mbox{N}_f)_{\mbox{\scriptsize R}}$ and is a singlet under
$\mbox{SU}(\mbox{N}_f)_{\mbox{\scriptsize L}}$. The $\mbox{N}_f^2$
operators $\bar{u}_{\mbox{\scriptsize R}} u_{\mbox{\scriptsize
L}},\;\bar{u}_{\mbox{\scriptsize R}} d_{\mbox{\scriptsize
L}},\;\ldots$
constitute the irreducible representation ($\mbox{\underline{N}}_f^\star ,
\mbox{\underline{N}}_f$)
of $\mbox{SU}(\mbox{N}_f)_{\mbox{\scriptsize R}} \times
\mbox{SU}(\mbox{N}_f)_{\mbox{\scriptsize L}}$, while their hermitean
conjugates transform according to
($\mbox{\underline{N}}_f ,\mbox{\underline{N}}_f^\star$). The
scalar $\bar{q}q = \bar{q}_{\mbox{\scriptsize R}} q_{\mbox{\scriptsize L}} +
\bar{q}_{\mbox{\scriptsize L}} q_{\mbox{\scriptsize R}}$
thus belongs to the direct sum of these two representations.

Since the dimension four operator
tr$_cG_{\mu \nu}G^{\mu \nu}$ is a singlet under G, its vacuum expectation
value, the gluon condensate, does not represent an order parameter.
At
dimension five or six, however, several H-invariant Lorentz scalars
may be built, which transform
in a nontrivial manner under G:
$\bar{q}\sigma_{\mu\nu}G^{\mu\nu}q$,
$(\bar{q}q)^2$, $(\bar{q}\gamma_5 q)^2$,
$(\bar{q}\lambda_aq)^2$, $(\bar{q}\gamma_\mu\lambda_a q)^2$,
$(\bar{q}\sigma_{\mu\nu} q)^2,\,\ldots\;$
Not all of these give rise to independent order parameters. The
expectation values $\lvac(\bar{q}q)^2\rvac$ and
$\lvac(\bar{q}\gamma_5 q)^2\rvac$, e.g.,
are different from zero, even if the state $\rvac$ is symmetric with
respect to G: these operators belong to a multiplet, whose
decomposition into irreducible representations
contains a singlet. In the difference
$(\bar{q}q)^2-(\bar{q}\gamma_5 q)^2$, however, the singlet drops out;
the expectation value $\lvac(\bar{q}q)^2-(\bar{q}\gamma_5 q)^2\rvac$ does
represent an
order parameter, which is independent of $\lvac\bar{q}q\rvac $ (for
a G-invariant state, $\lvac(\bar{q}q)^2\rvac=
\lvac(\bar{q}\gamma_5 q)^2\rvac$).
%%%%%%%%%%%%%%%%%%%%%%%%%%%%%%%%%%%%%%%%%%%%%%%%%%%%%%%%%%%%%%%%%%%%%
\section{Quark masses}
\label{masses}
The preceding discussion concerns the fictitious world where all of the
quark masses are set equal to zero. In reality, the Lagrangian of QCD
contains a quark mass term, which breaks chiral symmetry. The divergence of the
currents introduced above is determined by the quark mass matrix,
\bdm\begin{array}{ll}
\partial_\mu V_a^\mu=\mbox{$\frac{1}{2}$}i\bar{q}(m \lambda_a-\lambda_a
m) q\;\;,\;\;&
\partial_\mu V^\mu_0=0\;\;,\\
\partial_\mu A_a^\mu =\mbox{$\frac{1}{2}$}i \bar{q}(m\lambda_a+\lambda_a
m) \gamma_5 q\;\;,\;\;&
\partial_\mu A^\mu_0=2i\bar{q}m\gamma_5q+
\frac{\mbox{N}_f}{\mbox{8$\pi ^2$}} \mbox{tr}_c G_{\mu \nu} \tilde{G}^{\mu
\nu }\fs\end{array}\edm
Since the quark masses are different from one another, only
the diagonal vector currents are conserved:
$\bar{u} \gamma^\mu u$, $\bar{d} \gamma^\mu d, \ldots\,$
There is one conserved charge
for every one of the quark flavours. Baryon number,
electric charge, strangeness, charm, etc. are linear combinations thereof.
Accordingly, the symmetry
group of
real QCD is the subgroup generated by the diagonal vector currens, $\mbox{G}_1
= \mbox{U}(1)^{\mbox{\scriptsize N}_f}
\subset\mbox{G}_0$.

It so happens, however, that some of the quark masses are
small. One may treat these as perturbations --- QCD
possesses
an {\it approximate} chiral symmetry. If only the small quark masses are turned
off, the Lagrangian acquires a symmetry group G which is
is larger than $\mbox{G}_1$, but smaller than
the group of maximal symmetry arising if all of the quark
masses are set equal to zero,
$\mbox{G}_1\subset\mbox{G}\subset\mbox{G}_0$.
I now discuss the phenomenological evidence for the
occurrence of such an approximate symmetry.

A striking property of the observed pattern of bound states is that they
come in nearly degenerate {\it isospin} multiplets:
$(\pi^+,\pi^0,\,\pi^-),\,(K^+,\,K^0),$ $(\bar{K}^0,\,K^-)$, $(P,\,N),\,
(\Sigma^+,\,\Sigma^0,\,\Sigma^-),\;\ldots\;$
In fact, the splittings within these multiplets are so small that, for a long
time, isospin was assumed to represent an
{\it exact} symmetry of the strong interactions; the observed
small
mass difference between neutron and proton or $K^0$ and $K^+$ was blamed on the
electromagnetic interaction.

We now know that this picture is incorrect: the
bulk of isospin breaking does not originate in the electromagnetic
fields, which surround the various particles, but is due to the fact that
the $d$-quark is somewhat heavier than the $u$-quark: isospin only represents
an approximate symmetry of the strong interactions. The symmetry
arises in the theoretical limiting case, where $m_u=m_d$. In this limit,
the flavours $u$ and $d$
become
indistinguishable, as far as QCD is concerned, such that the
Lagrangian acquires an exact symmetry with respect to
\bdm \begin{array}{lll}u&\rightarrow&\alpha u +\beta d\\
d&\rightarrow&\gamma u + \delta d\end{array}\hspace{3em}
V=\left(\!\begin{array}{ll}\alpha&\!\!\beta\\
\gamma&\!\!\delta\end{array}\!\right)\;\in\; \mbox{SU(2)}\fs\edm
The group is generated by the charges of the three vector currents
\bdm V^\mu_+ =\bar{u}\gamma^\mu
d\;\;,\;\;V^\mu_0=\mbox{$\frac{1}{2}$}(\bar{u}\gamma^\mu u  -\bar{d}\gamma^\mu
d)\;\;,\;\;V^\mu_-=\bar{d}\gamma^\mu u\fs\edm
The transformation law states that $u$ and $d$ form an
isospin doublet, while the remaining
flavours $s,\,c,\,\ldots\,$ are singlets. In addition, the
U(1) charges associated with $\bar{u}\gamma^\mu u+\bar{d}\gamma^\mu
d,\;\bar{s}\gamma^\mu s,\;\ldots\,$ are also conserved, such that, in the limit
$m_u=m_d$, the full symmetry group of the Lagrangian is given by
$\mbox{G}=\mbox{SU}(2)_{\indV}
\times\mbox{U(1)}^{\,\mbox{\scriptsize N}_f-1}$.

We may exhibit the piece
of the QCD Hamiltonian which breaks isospin symmetry by rewriting the mass
term of the $u$ and $d$ quarks in the form
\bdm m_u\bar{u}u+m_d\bar{d}d=\mbox{$\frac{1}{2}$}(m_u+m_d)(\bar{u}u+
\bar{d}d)+\mbox{$\frac{1}{2}$}(m_u-m_d)(\bar{u}u-
\bar{d}d)\fs\edm
The remainder of the Hamiltonian is invariant
under isospin transformations and the same is true of the operator
$\bar{u}u+\bar{d}d$. The
QCD Hamiltonian thus consists of an isospin invariant part
$\stackrel{\rule{0.7em}{0.05em}}{\bf H }_0$ and a
symmetry breaking term, proportional to the mass difference $m_u-m_d$,
\bdm
{\bf H}_{\QCD}=\,\stackrel{\rule{0.7em}{0.05em}}{\bf H }_0
+\stackrel{\rule{0.7em}{0.05em}}{\bf H }_{1}\;\;,\;\;
\stackrel{\rule{0.7em}{0.05em}}{\bf H }_{1}\,=
\int\!d^3\!x\,\mbox{$\frac{1}{2}$}(m_u - m_d) (\bar{u}u - \bar{d}d)\fs\edm
The strength of isospin breaking is controlled by the quantity
$m_u-m_d$, which plays the role of a {\it symmetry breaking
parameter}. The fact that the multiplets are nearly degenerate
implies that the operator $\stackrel{\rule{0.7em}{0.05em}}{\bf H }_{1}$
only represents a small
perturbation: the mass difference $m_u-m_d$ must be very small.

On the basis of the few strange particles, which had been discovered in the
course of the 1950's,
Gell-Mann and Ne'eman \cite{Gell-Mann} inferred that the strong
interactions exhibit a further approximate symmetry,
of the same qualitative nature as
isospin, but more strongly broken. The symmetry,
termed the {\it eightfold way}, played a decisive role in unravelling
the quark degrees of freedom. By now, it has become evident that the
mesonic and baryonic levels are indeed grouped
in multiplets of SU(3) --- singlets, octets, decuplets --- and there is also
good phenomenological support for the corresponding symmetry relations among
the various observable quantities.

In the framework of QCD, eightfold way symmetry occurs
in the theoretical limit, where the three lightest quarks are
given the same mass, $m_u=m_d=m_s$. The Hamiltonian then becomes invariant
under the transformation
\bdm \left(\!\begin{array}{c} u\\d\\
s \end{array}\!\right) \rightarrow
V\left(\!\begin{array}{c} u\\d\\s \end{array}\!\right)\hspace{3em}V\in
\mbox{SU(3)}\edm
of the quarks fields. Again, the full symmetry group in addition contains
several U(1) factors, $\mbox{G}= \mbox{SU(3)}_{\indV}\times
\mbox{U(1)}^{\,\mbox{\scriptsize N}_f -2}$. In the limit $m_u=m_d=m_s$, the
spectrum of the theory consists of degenerate multiplets of this group. The
degeneracy is lifted by the mass differences
$m_s-m_d$ and $m_d-m_u$, which represent the symmetry breaking parameters in
this case.
Since the eightfold way does represent an approximate symmetry of the strong
interactions, both of these mass differences must be small. Moreover, the
observed level pattern requires
$\mid\!m_d-m_u\mid\,\ll\,\mid\!m_s-m_d\!\mid$.

Formally, the above discussion may be extended to include additional
flavours. One may even consider
the theoretical limit, where all of the $\mbox{N}_f$ quarks are given the same
mass.\footnote{The massless theory discussed in section \ref{qcdmq}
represents a
special case.}
The Hamiltonian then becomes invariant under
$\mbox{U}(\mbox{N}_f)_{\indV}$ and the spectrum consists of
degenerate multiplets
of this group. The extension, however, does not correspond to an
approximate symmetry. The lightest
pseudoscalar bound state with the quantum numbers of $\bar{d}c$, e.g., sits at
$M_{D^+}\simeq 1.87\,\mbox{GeV}$. If the mass of the charmed quark is
set equal to $m_u$, this state becomes degenerate with the $\pi^+$.
Clearly, the mass
difference $m_c-m_u$, which
plays the role of a symmetry breaking parameter in this case, does not
represent a small perturbation. We do not know why the quark masses
follow the pattern observed in nature, nor do we understand the
equally queer pattern of lepton masses. It so happens that the mass
differences between $u,\,d$ and,$s$ are small, such that the
eightfold
way represents a decent approximate symmetry. The remaining masses
turn out to be quite
different, so that the mesons and baryons which contain them do not
closely resemble
the corresponding light quark bound states. A
more promising line of approach is {\it heavy quark symmetry}, which analyzes
the properties of their bound states by treating $c,\,b,\,t$ as
infinitely heavy \cite{Isgur}.
%%%%%%%%%%%%%%%%%%%%%%%%%%%%%%%%%%%%%%%%%%%%%%%%%%%%%%%%%%%%%%%%%%%%%%
\section{Approximate chiral symmetry}
The approximate symmetries discussed in the preceding section explain
why the bound states of QCD exhibit a multiplet pattern,
but they do not account
for an observation which is equally striking and which plays a crucial role in
strong interaction physics: the mass gap of the theory, $M_\pi,\,$ is
remarkably small.
The approximate symmetry which hides behind this observation was discovered by
Nambu \cite{Nambu}. It originates in the fact
that $m_u$ and $m_d$ happen to
be small.

Consider first the limit, where a single one of the quarks is taken massless,
$m_u=0$, while the remaining masses are kept fixed at their physical values.
The QCD Lagrangian then becomes invariant under the chiral transformation
$u\rightarrow \exp (i\beta\gamma_5) u$. As mentioned above, this
symmetry of the
Lagrangian, however, is ruined by the U(1) anomaly --- the divergence of the
Noether current $\bar{u}\gamma^\mu\gamma_5 u$ is different from zero. So, the
limit $m_u\rightarrow 0$ does not give rise to a higher degree of symmetry: the
symmetry group is the same as for $m_u\neq 0$.

The theory does acquire more symmetry
if two of the quark masses are taken equal, as it then becomes invariant under
isospin rotations. If $m_u$ and $m_d$
are not only taken the same, but are put equal to zero, the symmetry group
is increased further. The Lagrangian then
becomes invariant with respect to a set of {\it chiral} transformations:
independent
isospin rotations of the right- and lefthanded components of $u$ and $d$,
\bdm\left(\!\!\begin{array}{c} u_{\indR}\\d_{\indR}
\end{array}\!\!\right)\rightarrow V_{\indR}
\left(\!\!\begin{array}{c} u_{\indR}\\d_{\indR}
\end{array}\!\!\right)\;\;,\;\;
\left(\!\!\begin{array}{c} u_{\indL}\\d_{\indL}
\end{array}\!\!\right)\rightarrow V_{\indL}
\left(\!\!\begin{array}{c} u_{\indL}\\d_{\indL}
\end{array}\!\!\right)\;\;\;\;\;\;\;V_{\indR},\,
V_{\indL} \in \mbox{SU(2)}\fs\edm
In contrast to the chiral U(1) transformation considered above, this symmetry
of the classical Lagrangian does
survive quantization: in the limit $m_u=m_d=0$,
the theory becomes invariant with respect to the
group $\mbox{G}=\mbox{SU(2)}_{\indR}\times
\mbox{SU(2)}_{\indL}\times
\mbox{U(1)}^{\mbox{\scriptsize N}_f-1}.$ As discussed in section \ref{qcdmq},
chiral symmetry
is broken spontaneously, the ground state being invariant only under the
subgroup generated by the charges of
the vector currents. For the limit under consideration, where only two of the
quark flavours are taken massless, the ground state is symmetric under
$\mbox{H}=\mbox{SU(2)}_{\indV} \times \mbox{U(1)}^{\mbox{\scriptsize N}_f-1}$,
such that G/H = SU(2). Accordingly, the spectrum contains three Goldstone
bosons, with the quantum numbers of $\pi^+,\pi^0,\pi^-$.

In reality,
chiral symmetry is broken, not only spontaneously, but also explicitly, by the
quark masses. As above, we may split the Hamiltonian into a
piece which is invariant under the symmetry group of interest and a piece
which breaks the symmetry. In the present case, the symmetry breaking part
is the mass term of the $u$ and $d$ quarks,
\bdm {\bf H}_{\QCD}={\bf H}_0^\prime+
{\bf H}_{sb}^\prime\;\;,\;\; {\bf H}_{sb}^\prime
=\int\!d^3\!x(m_u\bar{u}u+m_d\bar{d}d)\fs\edm
If the symmetry breaking parameters $m_u$ and $m_d$ are small, the
spectrum of ${\bf H}_{\QCD}$ must be close to the spectrum of
${\bf H}_0^\prime$. In
particular, it must
contain three one-particle states, whose mass tends to zero when the symmetry
breaking is turned off. In fact, we will see that the pion mass
tends to zero in proportion to the square root of $m_u+m_d$,
\bdm M_{\pi}^2\propto (m_u+m_d)\fs\edm
The observed spectrum is remarkably close to the
one which would result if chiral symmetry was an exact symmetry of the strong
interactions:
the pions are by
far the lightest hadrons.
So, the fact that the
pions are light can be understood: they are the Goldstone bosons of a
spontaneously broken approximate symmetry.

Now, we already noted that the difference between $m_s$ and $m_u$ or $m_d$
must be small, because the eightfold way represents an approximate symmetry
of the strong interactions. Since the smallness of the pion mass requires
$m_u$ and $m_d$ to be small, we conclude that
all three quarks must be light --- the
mass terms of $u,\,d$ and $s$ only represent a
perturbation. Moreover, the inequality $\mid\!m_d-m_u\!\mid\ll
\mid\!m_s-m_d\!\mid$, which
follows from the fact that isospin breaking is much smaller than the breaking
of eightfoldway symmetry, implies $m_u,\,m_d\ll m_s$.

The Hamiltonian may be decomposed according to
\bdm {\bf H}_{\QCD}={\bf H}_0+ {\bf H}_{sb}\;\;\;,\;\;\;
{\bf H}_{sb}
=\int\!d^3\!x(m_u\bar{u}u+m_d\bar{d}d+m_s\bar{s}s)\fs\edm
The first term, ${\bf H}_0$,
describes three massless flavours $(u,\,d,\,s)$ as well
as three massive ones $(c,\,b,\,t)$ and is symmetric with respect to the
group $\mbox{G}=\mbox{SU(3)}_{\indR}\times
\mbox{SU(3)}_{\indL}\times
\mbox{U(1)}^{\mbox{\scriptsize N}_f-2}$. The second term,
${\bf H}_{sb}$,
breaks this symmetry to the subgroup $\mbox{H}=
\mbox{U(1)}^{\mbox{\scriptsize N}_f}$. Since the symmetry breaking parameters
$m_u,\,m_d$ and $m_s$ are small, the properties of the theory can be analyzed
by treating ${\bf H}_{sb}$ as a perturbation.
The corresponding
perturbation series amounts to an expansion in powers of $m_u,\,m_d$ and $m_s$.
The point here is that the two groups
$\mbox{SU(2)}_{\indR}\times\mbox{SU(2)}_{\indL}$
and $\mbox{SU(3)}_{\indV}$ can be approximate symmetries of the
Hamiltonian only if the group
$\mbox{SU(3)}_{\indR}\times\mbox{SU(3)}_{\indL}$
represents an approximate symmetry, too.

There is an immediate experimental check: the eight lightest bound states,
$\pi^+,\,\pi^0,\,\pi^-,\,K^+,\,K^0,\,\bar{K}^0,\,K^-,\,\eta,\, $ indeed
carry precisely the quantum numbers of the Goldstone bosons
generated by the spontaneous breakdown
$\mbox{SU(3)}_{\indR}\times\mbox{SU(3)}_{\indL}
\rightarrow \mbox{SU(3)}_{\indV}$. They are not massless,
because the quark masses break the symmetry, but the breaking is small enough
for these levels to remain lowest.

The above arguments rely on two
phenomenological observations:
\begin{description}
\item{(a)}
The pion mass is small compared to the masses of all other hadrons.
This indicates that the strong interactions possess an approximate,
spontaneously broken symmetry, with the pions as the corresponding
Goldstone bosons. Indeed, the Lagrangian of QCD exhibits an approximate
symmetry with the proper quantum numbers, provided both $m_u$ and $m_d$
are small.

\item{(b)}
The multiplet structure seen in the particle data tables
indicates that SU(3) is an approximate symmetry of the
strong interactions. For QCD to exhibit such a symmetry, the
mass differences $m_d-m_u$ and $m_s-m_d$ must be small. Together with (a),
this implies that all three quarks $u,d$ and $s$ are light, such that
the Lagrangian of QCD exhibits an approximate chiral
symmetry, with G = SU(3)$_{\indR}\times$SU(3)$_{\indL}$.

\end{description}

The masses of the other quarks occurring in the Standard Model, on the other
hand, cannot be treated as a perturbation. Since the corresponding quark fields
$c(x),\,b(x)$ and $t(x)$ are invariant under G, their contribution to the
Lagrangian may be included in the SU(3)$_{\indR}
\times$SU(3)$_{\indL}$ invariant part of the
Hamiltonian, ${\bf H}_0$, and does not significantly affect the
following discussion. Conversely, that analysis will not shed any light
on the properties of bound states involving heavy quarks.
%%%%%%%%%%%%%%%%%%%%%%%%%%%%%%%%%%%%%%%%%%%%%%%%%%%%%%%%%%%%%%%%%%%%%
\section{Pion pole dominance}
\label{ppd}

The
exchange of Goldstone bosons gives rise to singularities in the low energy
region, in particular, to
poles, connected with one-particle reducible
contributions. The analysis of these singularities
involves an
assumption, referred to as the PCAC or {\it pion pole dominance hypothesis} :
one postulates that, at sufficiently small momenta, the
one-particle-singularities dominate over the remainder.

To discuss the content of this hypothesis, I return to the case of a
spontaneously broken {\it exact} symmetry. For definiteness, I consider QCD
with two massless flavours, $m_u=m_d=0$. The spectrum then
contains three massless pseudoscalars, $M_{\pi^+}=M_{\pi^0}=M_{\pi^-}=0$.
In the Green functions of the theory, massless
one-particle states manifest themselves as poles at $p^2=0$. The
two-point function of the axial current, e.g., contains such a pole, arising
from the exchange of a pion between the two currents. Current
conservation and isospin invariance imply that this particular Green function
is of the form
\bdm \int\!d^4\!xe^{ipx}\lvac TA^\mu_a(x)A^\nu_b(0)\rvac
=i\delta_{ab}(p^\mu p^\nu-g^{\mu \nu}p^2)\Pi(p^2)\fs\edm
The pole term arises from those matrix elements of the current which
connect the vacuum to the one-pion states. On account of Lorentz invariance
and isospin symmetry, these are of the form
\be
\label{qcd12}
\lvac A^\mu_a | \pi^b(p)\rangle = i p^\mu \,\delta_a^b \,F\fs
\ee
The phase of
the states $|\pi^a(p)\rangle$ may be chosen such that the constant $F$ is
real and positive. The corresponding contribution to the two-point function
is proportional to $F^2$,
\bdm
\label{eft19}
\Pi (p^2) = \frac{F^2}{-p^2-i\epsilon} +\stackrel{\rule{0.6em}{0.03em}\,}{\Pi}
\!(p^2)\fs \edm
In this example, the pion pole dominance hypothesis boils down to
the assumption that, at low momenta, the pole term due to one-pion
exchange dominates over the remainder,
which contains branch points due to multipion exchange, as
well as singularities associated with the exchange of massive
particles. Accordingly, the low energy behaviour of the two-point function is
determined by the constant $F$;
the remainder, $\stackrel{\rule{0.6em}{0.03em}\,}{\Pi}\!(p^2)$,
only contributes if the low energy expansion is
carried beyond leading order.

The constant $F$ plays a central role in the low energy analysis. It specifies
the vacuum-to-pion
matrix element of the axial current and represents
the overlap of the states $|\pi^a(p)\rangle$ with those obtained by applying
the axial charges to the vacuum.\footnote{Note that the scalar product of the
two states is meaningful only
in the presence of an infrared cutoff; one may, e.g., replace the charge by
$\int\!d^3\!x f(x)A^0_a(x)$, where the test function $f(x)$ is equal to
one on some finite region of space, but vanishes at large distances.} The
Goldstone theorem asserts that $F$ is different from zero
\cite{Goldstone}.

Since the matrix element $\lvac A^\mu_a | \pi^b(p)\rangle\, $ determines
the rate of the weak decay $\pi
\rightarrow \mu \nu $, the quantity $F$ is referred to as the pion decay
constant. The measured decay rate implies
$F\simeq 93\, \mbox{MeV}$. Note that the magnitude of the matrix element
(\ref{qcd12}) changes if the quark masses are varied. As we are
presently considering
the theoretical limit $m_u=m_d=0$,
we cannot
use the experimental information as such, but have to distinguish
between the physical value of the pion decay constant and the value which
results if the quarks are taken massless. The difference is
characteristic of the chiral symmetry breaking effects
to be discussed later on.

The pole dominance hypothesis may be extended to any other matrix element,
which
receives one-particle reducible contributions from pion exchange. These are
described by a product of pole factors (one for each of the exchanged
pions) and a residue, representing a product of one-particle-{\it irreducible}
matrix elements. The pion pole dominance hypothesis is the assumption that,
at small momenta, the pole terms
dominate over the remainder. The four-point
function of the current, e.g., contains a contribution involving the emission
or absorption
of a pion by each one of the four currents. The residue is the product of
a one-particle irreducible amplitude, describing the
interaction among the four pions, and four matrix elements of the
type $\lvac A^\mu|\pi\rangle$, representing the interaction of the pions with
the current. Contributions from the
exchange of less than four pions also occur. According to pion pole
dominance, these pole terms dominate the four-point function at low
momenta.

In the case of the function $\Pi(p^2)$, the
residue of the pole is a constant, for kinematical
reasons. In general, however, the residue still represents a
function of the momenta, which contains singularites and does not admit a
straightforward
Taylor series expansion. In the four-point function, e.g., the residue is
proportional to the elastic scattering amplitude, which
contains branch point singularities related to rescattering processes
with two or more pions as intermediate states.
Both in the "Current algebra and
PCAC" approach and in the effective Lagrangian method, one assumes that
singularites associated with multipion exchange only occur at subleading
orders of the low energy expansion; retaining only the leading term of the
expansion, the residues reduce to polynomials of the momenta. The coefficients
occurring therein play a role analogous to the leading Taylor coefficients of
the low energy expansion for theories with an energy gap.
%%%%%%%%%%%%%%%%%%%%%%%%%%%%%%%%%%%%%%%%%%%%%%%%%%%%%%%%%%%%%%%%%%%%%%%%%
\section{Strength of the effective interaction}
\label{strength}
As an immediate application of the pion pole dominance hypothesis, I now show
that, at low energies, the hidden symmetry prevents the Goldstone bosons from
interacting with one another. This property is essential for the
consistency of $\chi$PT.

For simplicity, I again consider two massless
flavours, such
that the components $a=1,2,3$ of the axial current $A^\mu_a(x)$ are strictly
conserved. The argument relies on the properties of the the
probability amplitude for the currents to create pions
out of the vacuum. Current conservation requires this amplitude to obey the
condition \be
\label{eft30}
p_\mu\;\langle \pi^{a_1}(p_1),\pi^{a_2}(p_2),\ldots \;\mbox{out}|
A^\mu_a\rvac=0\co \ee
where $p^\mu= p^\mu_1+p^\mu_2+\ldots $ is the four-momentum of the final
state.

The probability amplitude for the occurrence of a single pion is linear in the
momentum,
\be
\label{eft31}
\lvac
A^\mu_b|\pi^a(p)\rangle\;=\;
\langle\pi^a(p)| A^\mu_b\rvac^\star \;=\;
 i \,p_\mu \,\delta^a_b\,F \ee
and obeys the conservation law,
provided the pions are massless, $p^2=0$.
The amplitudes for the creation of several particles, however, may contain
singularities due to pion exchange. The graph\begin{figure}
\begin{center}
\mbox{\epsfxsize=13.5cm
      \epsfbox{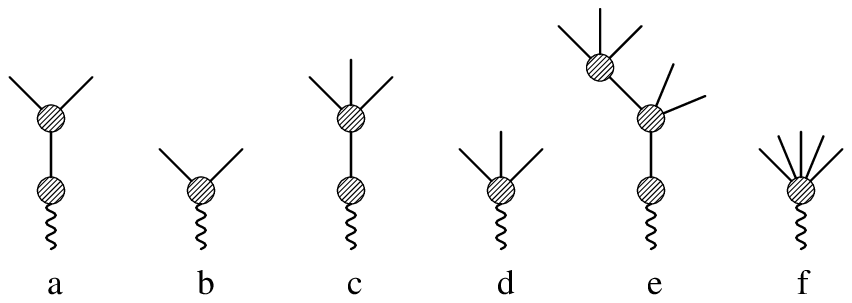} }
\end{center}
\caption{Pion production
by an external current. The internal lines represent
pion poles, the external ones stand for on-shell pions. The current is
denoted by a wiggly line and the shaded circles indicate one-particle
irreducible amplitudes.} \label{fig1}\end{figure}\addtocounter{figuren}{1}
shown in figure \ref{fig1}a, e.g., represents a one-particle reducible
contribution to the production amplitude for two pions, in fact the only such
contribution which can occur in an amplitude with three legs. Describing the
three-pion-vertex by the function $g_{a_1a_2a_3}(p_1,p_2,p_3)$, this
graph yields the term \bdm
\label{eft32}
\langle\pi^{a_1}(p_1),\pi^{a_2}(p_2) \;\mbox{out}| A^\mu_{a_3}\rvac
=i
\frac{p_3^\mu \,F}{p_3^2+i\epsilon}\;g_{a_1a_2a_3}(p_1,p_2,p_3)+
\ldots\edm
with $p_1+p_2+p_3=0$. According to the preceding section, the
leading term of the low energy expansion of the residue is a
polynomial in the momenta. Suppose the expansion of the vertex
starts with a constant term, such that, at leading order of the expansion, the
vertex is replaced by a set of coupling constants $g_{a_1a_2a_3}(0,0,0)\neq 0$.
The contribution of graph \ref{fig1}a is of then of order
$p^{-1}$. Since this graph represents the only one-particle reducible
contribution, the remainder (fig.\ref{fig1}b)
is free of poles; at leading
order of the low
energy expansion, the remainder at most amounts to a constant term and
cannot compete with graph \ref{fig1}a. The pion
pole dominance hypothesis thus
implies that the leading contribution to the conservation law (\ref{eft30}) is
given by
\bdm
\label{eft33}
p_\mu\;\langle \pi^{a_1}(p_1),\pi^{a_2}(p_2) \;\mbox{out}| A^\mu_{a_3}\rvac
=i F\;g_{a_1a_2a_3}(0,0,0)+O(p)\co
\edm
where the symbol $O(p)$ indicates that the terms omitted involve at least
one power of momentum and hence vanish if all momenta are sent to
zero. We thus arrive at the low energy theorem
\bdm
\label{eft34a}
g_{a_1a_2a_3}(0,0,0)=0\fs
\edm

Actually, the above argument kills a dead fly: a triple pion vertex does not
occur in any case, because it would violate parity as well as G-parity.
The reason for considering this vertex, nevertheless, is that the argument
given immediately
generalizes to vertices involving any number of pions. Consider, e.g., the
amplitude for the
production of a final state with three pions. In the
absence of a three-pion vertex, this amplitude again contains a single
one-particle reducible contribution, shown in figure \ref{fig1}c. The
contribution is of the form
\bdm
\label{eft35}
\langle\pi^{a_1}(p_1),\pi^{a_2}(p_2)\pi^{a_3}(p_3) \;\mbox{out}\mid\!
A^\mu_{a_4}\rvac =i
\frac{p^\mu_4 \,F}{p^2_4+i\epsilon}\;g_{a_1a_2a_3a_4}
(p_1,p_2,p_3,p_4) + \ldots\edm
At low momenta, this term again dominates over the remainder
(graph \ref{fig1}d), such that current conservation implies
\bdm
\label{eft36}
g_{a_1a_2a_3a_4}(0,0,0,0)=0\fs
\edm
In other words, the hidden symmetry prevents pions of zero momentum from
scattering elastically.

The production amplitudes for five or more
pions contain multiple
poles. The one-particle reducible graph shown in figure \ref{fig1}e, e.g.,
involves a double pole from the two internal lines. Since each of the
four-pion-vertices occurring there, however, is suppressed by two powers
of momentum, the contribution generated by this graph is only of $O(p)$
and does, therefore, not show up when evaluating the conservation law
(\ref{eft30})
in the zero momentum limit. Again, current conservation can only be satisfied
if the six-pion-vertex (graph \ref{fig1}f) vanishes at zero
momentum. By induction, the argument extends to vertices involving any number
of pions: current conservation implies that all of these vertices are of
order $p^2$ and disappear if the momenta
of the pions tend to zero. Accordingly, the low energy expansion of the
production amplitudes $\langle\pi\pi\ldots\;\mbox{out} | A^\mu\rvac$
only starts at $O(p)$, irrespective of the number of pions produced.
The scattering amplitudes are described by the same graphs, except that some of
the pions must be crossed from the final to the initial state and the internal
line linking these to the current is to be replaced by an external
line. This shows that, independently of the number of pions occurring in
the initial and final states, the scattering amplitudes are at most of
$O(p^2)$.

At low energies, the interaction among the
Goldstone bosons thus becomes weak --- pions of zero energy do not interact at
all. This is in marked
contrast to the interaction among the quarks and gluons, which is strong at low
energies, because QCD is an asymptotically free, infrared enslaved theory.
The qualitative difference is crucial for chiral perturbation theory to be
coherent: in this framework, the interaction among the
Goldstone bosons is treated as a perturbation. The opposite behaviour in the
underlying theory prevents a perturbative low energy analysis of the
interaction among the quarks and gluons.
%%%%%%%%%%%%%%%%%%%%%%%%%%%%%%%%%%%%%%%%%%%%%%%%%%%%%%%%%%%%%%%%%%%%%%
\section{Effective Lagrangian}
\label{lag}
The effective Lagrangian method is based on the following idea.
The graphs shown in figure \ref{fig1}
may be viewed
as tree graphs of a field theory, which involves {\it pion fields} as basic
variables. Since the Goldstone bosons do not carry spin, they are described by
scalar fields, which I denote by $\pi^a(x)$.
The fields are in one-to-one correspondence with the massless
one-particle-states $\mid\!\pi^a(p)\rangle$ occurring in the spectrum of the
theory.

In this language, the pole terms generated by pion exchange arise from
the propagation of the pion field, described by the correlation function
\bea
\label{eff1}
\lvac T\,\pi^a(x) \pi^b (y)\rvac
=
\frac{1}{i}\delta^{ab}\Delta_0(x-y)\no
\Delta_0(z)=
\int\! \frac{d^4\!p}{(2\pi )^4}\;\frac{e^{-ipz}}{-p^2-i\epsilon}=
 \frac{1}{4\pi^2i(z^2-i\epsilon)}\fs\nonumber\eea
The Feynman propagator $\Delta_0(x-y)$ represents the transition amplitude for
a pion emitted at the point
$x$ to reach the point $y$, or vice versa: the
propagator is an even function, $\Delta_0(x-y)=\Delta_0(y-x)$. Since the
Fourier transform
thereof is given by $1/(-p^2-i\epsilon)$, the propagation of the field
between the various vertices indeed yields
the relevant pole terms occurring in one-particle reducible graphs.

The propagator is determined by the kinetic part of the Lagrangian and vice
versa. For the effective field theory to yield the massless scalar
propagator, the kinetic term must be identified with the
standard expression describing scalar free fields,
\bdm {\cal L}_{kin}=\mbox{$\frac{1}{2}$}\partial_\mu
\pi^a\partial^\mu\pi^a\fs\edm

The vertices, on the other hand, represent interactions of the
field. Some of the vertices of figure \ref{fig1}
describe emission or absorption of pions by the currents, others
exclusively join pion lines.
In the language of the effective
field theory, the purely pionic vertices correspond to terms in the interaction
Lagrangian,
\bdm
\label{eff5}
{\cal L}_{int}={\cal L}_{int}(\pi ,\partial \pi ,\partial^2\pi,\,\ldots )\fs
\edm

A momentum independent vertex joining four pion lines, e.g., corresponds to an
interaction term of the form $g_{abcd}\,\pi^a\pi^b\pi^c\pi^d$, while a
term of the type $g^{\,\prime}_{abcd}\,\partial_\mu \pi^a\partial^\mu
\pi^b\pi^c\pi^d$ generates a vertex with two powers of momentum.
The translation of the various vertices into
corresponding terms of the interaction Lagrangian is trivial: if the vertex
in question joins $P$ pion lines and involves a polynomial in the momenta of
degree $D$, the corresponding term in ${\cal L}_{int}$ contains $P$
pion fields and $D$ derivatives. Since an interaction involves at least three
pions, $P\geq 3$. Moreover, Lorentz invariance implies
that $D$ is even, such that the derivative expansion of the interaction
Lagrangian starts with
\be
\label{eff6r}
{\cal L}_{int}= g^0(\pi) +
g^1_{\,ab}(\pi)\partial_\mu\pi^a\partial^\mu\pi^b+g^2_{\;a}(\pi)\,
\raisebox{0.2em}{\fbox{\rule[0.15em]{0em}{0em}\,}}\,\pi^a+  O(p^4)\co
\ee
where the omitted terms involve four or more derivatives.
The Taylor series of the function $g^0(\pi)$ in powers of $\pi$ yields all
vertices which are momentum independent:
\bdm
\label{eff6s}
g^0(\pi)=
\mbox{$\frac{1}{3!}$}\,g^{\,0}_{abc}\pi^a\pi^b\pi^c
+\mbox{$\frac{1}{4!}$}\,g^{\,0}_{abcd}\pi^a\pi^b\pi^c\pi^d +\ldots
\edm
Similarly, the expansion of the functions
$g^1_{\,ab}(\pi)$ and $g^2_{\;a}(\pi)$ in powers of $\pi$ generates all those
vertices, which contain two powers of momentum, etc.
The Lagrangian of the effective field theory merely collects the
information about the various vertices --- no more, no less.

The virtue of the representation in terms of effective fields is that the
tree graphs of a local field theory automatically obey the {\it cluster
decomposition property} : whenever a given number of pions meet, the same
vertex occurs, irrespective of the remainder of the diagram.
The presence of an interaction among four
pions, e.g., also manifests
itself in the process $\pi\pi\rightarrow\pi\pi\pi\pi$, through a tree graph
contribution containing two four-pion-vertices and one internal line,
which represents
the exchange of a pion between the two vertices (compare
figure \ref{fig1}e). The tree
graphs generated by the various interaction terms contained in the effective
Lagrangian automatically include these contributions.
Note that, at this
stage, only
the tree graphs of the effective field theory are relevant.

As shown in section \ref{strength},
current conservation not only implies that the Goldstone bosons are
massless, but
requires all of the vertices to vanish at zero momentum.
Hence, the effective Lagrangian does not contain any interaction terms
without derivatives, i.e.,
\be
\label{eff6t}
g^0(\pi)\equiv0\fs
\ee
In the language of effective field theory, the Goldstone theorem states
that the Lagrangian does not contain a mass term: a contribution $\propto
\pi^2$ does not occur. The relation
(\ref{eff6t}) may be viewed as a generalization of the theorem ---
it states that terms without derivatives are absent altogether.

The derivative expansion of the effective Lagrangian only starts
at $O(p^2)$ and contains terms with $2,4,6,\ldots$ derivatives,
\bdm\label{effla}
{\cal L}_{\eff} ={\cal L}_{\eff}^2+{\cal L}_{\eff}^4+
{\cal L}_{\eff}^6+\ldots\edm
As indicated in (\ref{eff6r}), Lorentz invariance permits two
different interaction terms of second order in the momenta. The second one
may, however, be rewritten as
$\partial_\mu\{g^2_{\;a}(\pi)\partial^\mu\pi^a\}-
\partial_{\raisebox{-0.12em}{$\scriptstyle b$}}\hspace{0.05em} g^2_{\;a}(\pi)
\partial_\mu\pi^a\partial^\mu\pi^b$.
Conservation of energy and momentum at each one of the vertices implies
that total derivatives do not contribute and the remainder may be absorbed
in $g^1_{\,ab}(\pi)$.
Without loss of generality, we may therefore set $g^2_{\;a}(\pi)=0$.
The leading contribution in
the derivative expansion of the effective Lagrangian then takes
the form
\bdm
\label{eff6x}
{\cal L}_{\eff}^2=
\mbox{$\frac{1}{2}$}g_{ab}(\pi)\partial_\mu\pi^a\partial^\mu\pi^b \co\edm
where I have amalgamated the kinetic term with the interaction, setting
$g_{ab}(\pi)\equiv \delta_{ab} + 2 g^1_{\,ab}(\pi)$. This shows that, at
leading order of the low energy expansion, the properties of the interaction
are characterized by
the function $g_{ab}(\pi)$. The expansion of this function
in powers of $\pi$,
\bdm\label{eff6z}
g_{ab}(\pi)=\delta_{ab}+\partial_c g_{ab}(0)\pi^c+\mbox{$\frac{1}{2}$}
\partial_{cd}g_{ab}(0)\pi^c\pi^d+\ldots\edm
generates all of the vertices of order $p^2$. The first term represents the
kinetic energy, the second generates the vertices with three
pion legs, the third specifies the interactions among four pions etc.
%%%%%%%%%%%%%%%%%%%%%%%%%%%%%%%%%%%%%%%%%%%%%%%%%%%%%%%%%%%%%%%%%%%%%
\section{Symmetries of the effective theory}
\label{sym}
The vertex shown in fig.\ref{fig1}a links the pion field to the axial
current.
In the famework of the effective description, this vertex corresponds to a
term
linear in the field,
\be\label{effcurrent}
A^\mu_a = - F \partial^\mu \pi^a + \ldots
\ee
while a term of the form $\partial_\mu \pi \pi \pi$, e.g., corresponds to a
vertex, where the axial current emits three pion lines.
The full current $A^\mu_a=A^\mu_a (\pi,\partial\pi,\ldots)$
consists of an infinite string of terms
(G-parity allows an arbitrary odd
number of pion fields and Lorentz invariance implies that
the number of derivatives is odd). The representation of the
vector current in terms of effective
fields consists of an analogous string (except that G-parity now only
permits terms built with an even number of fields).

The most important property of the
currents is that they are conserved,
$\partial_\mu V^\mu_a=\partial_\mu A^\mu_a=0$. The conservation law expresses
the fact that QCD with two massless flavours is invariant under
$\mbox{G}=\mbox{SU(2)}\times \mbox{SU(2)}$. The effective representation of
the currents must obey the same conservation law, i.e., the effective
theory must inherit the symmetries of the underlying one.
Indeed, if the effective Lagrangian is invariant under G, the Noether theorem
automatically provides a
representation for the vector and axial currents in terms of
the pion field and insures
that these currents are conserved. Note, however, that we are using the
Noether theorem in the wrong direction here: an invariant Lagrangian
leads to conserved currents, but the converse is not true. What counts is the
{\it action}; under the transformations generated by the symmetry group, the
Lagrangian may pick up a total derivative, such that the action is not
affected. This is precisely what happens in the presence of
anomalies. A similar phenomenon also arises in the nonrelativistic
domain: the effective Lagrangian describing the magnons of a ferromagnet
is invariant under rotations of the spin directions only up to a total
derivative \cite{Nonrel}.
These examples demonstrate that the effective Lagrangian is not
necessarily invariant under the global symmetry group G.
In fact, global symmetry does not fully determine the
structure of the effective Lagrangian.

A critical reader may also have doubts about the validity of the Noether
theorem in the present context, because its derivation
makes use of the equation of motion for the field. In the Feynman path
integral representation of the effective field theory, the
pion field freely fluctuates --- it is the variable of integration and does not
obey an equation of motion. In the tree approximation, the
theorem does hold, because
the tree graphs describe the classical limit.
Beyond the tree
approximation, however, the equation of motion is modified by the quantum
fluctuations of the field and a proper formulation of the Noether theorem
then becomes are rather subtle affair.

The symmetry properties of relativistic effective Lagrangians
are analyzed in detail in
\cite{found}. While the above discussion only concerns {\it global}
(i.e. space-time independent) symmetry operations, that analysis
relies on the Ward identities, which express the
symmetry
properties of the underlying theory in {\it local} form (see section
\ref{ward}). The main result established there is an {\it
invariance theorem}, valid for theories with
a Lorentz invariant ground state. The theorem
states that \begin{enumerate} \item[(i)] If the Ward identities do not
contain anomalous contributions, the effective Lagrangian is invariant
under G.
\item[(ii)] In the presence of anomalies, the effective Lagrangian
contains a
Wess-Zumino term at order $p^4$, whose form is known explicitly; the remainder
is invariant under G.
\end{enumerate}
For the proof, I refer to \cite{found}. The invariance theorem puts the
effective
Lagrangian method on firm footing. It demonstrates that this method is
strictly equivalent to the current algebra plus PCAC approach.
In either
case, the essential ingredient is the pion pole dominance hypothesis formulated
in section \ref{ppd}. As discussed there, this hypothesis fixes the leading
terms in the low energy expansion of the various Green functions up to
the Taylor coefficients. The Ward identities imply that the Taylor coefficients
occurring in different Green functions are related to one another. In the
current algebra plus PCAC analysis, one explicitly works out these
relations,
by investigating individual Green functions. The simplicity of the effective
Lagrangian method derives from the fact that
the Ward identities are equivalent to the simple statement that ${\cal
L}_{\eff}$ possesses the same local symmetry properties as the
Lagrangian of the underlying theory. This property allows one to explicitly
solve the constraints among the Taylor coefficients at
a given order of the low energy expansion, for all of the Green
functions at once.

In the preceding section, we focussed on the leading term in the derivative
expansion of the effective Lagrangian, ${\cal L}^2_{\eff}$. The
invariance theorem asserts that this term is invariant under G,
irrespective of anomalies ---
the Wess-Zumino term represents a specific contribution to ${\cal
L}^4_{\eff}$ and thus only matters if the low energy expansion is investigated
beyond leading order.
Let us now work out the consequences for the form of ${\cal L}^2_{\eff}$.
%%%%%%%%%%%%%%%%%%%%%%%%%%%%%%%%%%%%%%%%%%%%%%%%%%%%%%%%%%%%%
\section{Transformation law of the pion field}
\label{tp}
In the underlying theory, the group
acts on the quark fields, according to
\be\label{effsy2}
q_{\indR}
\stackrel{g}{\rightarrow}
V_{\indR} \,q_{\indR}\hspace{2em}
q_{\indL}\stackrel{g}{
\rightarrow} V_{\indL}\, q_{\indL}\fs\ee
In the
framework of the effective field theory, G instead acts on the pion fields,
through a representation of the form
\bdm\label{effsy3}
\pi^a\stackrel{g}{\rightarrow}\varphi^a(g,\pi)\fs\edm
The explicit expression for the corresponding Noether currents
is determined by the form of the function
$\varphi^a(g,\pi)$.
Remarkably, the representation property
\be\label{sym1} \varphi(g_1, \varphi (g_2, \pi)) =\varphi (g_1 g_2,
\pi)\ee fixes the mapping essentially uniquely \cite{Callan}.

To verify this claim, let us first consider
the
image of the origin, $\varphi (g,0)$. The composition law shows that the set of
elements $h$ which map the origin onto itself forms a subgroup $\mbox{H}
\subset \mbox{G}$.
Moreover, $\varphi (gh,0)$ coincides with $\varphi(g,0)$ for any
$g \in \mbox{G}$, $h
\in \mbox{H}$. Hence the function $\varphi (g,0)$ lives on the space G/H,
obtained from G by identifying elements $g,g'$ differing only by right
multiplication with a member of H, $g' = gh$. The function $\varphi(g,0)$
thus maps the elements of G/H into the space of pion field variables. The
mapping is invertible, because $\varphi (g_1,0) = \varphi(g_2,0)$ implies
$g^{-1}_1 g_2 \in \mbox{H}$. Geometically, the pion field variables $\pi^a$
may therefore be viewed as the coordinates of the quotient space G/H:
the Goldstone bosons live in this space.

Next,
choose a representative element $n$ in each one of the equivalence classes $\{
gh, h \in \mbox{H} \}$, such that every group element may uniquely be
decomposed as $g= nh$.
The composition law (\ref{sym1}) then shows that the image $n'$ of the element
$n$ under the action of $g \in \mbox{G}$ is obtained by decomposing the product
$gn$ into
$n'h$ --- the standard action of G on the space G/H. This implies that the
geometry fully fixes the transformation law
of the pion field, except for the freedom in the choice of coordinates on the
manifold G/H.

In the case of $\mbox{G} = \mbox{SU}(2) \times \mbox{SU}(2)$ and $\mbox{H}=
\mbox{SU}(2)$, the quotient G/H is the group SU(2). The pion field may be
represented as an element of this group, i.e., as
a $2 \times2$ matrix field $U(x)\! \in\! \mbox{SU}(2)$. Alternatively, we may
identify the pion field with the the three coordinates $\pi^1,\pi^2,\pi^3$,
needed to parametrize the group SU(2).
The choice of coordinates is not unique. Using canonical
coordinates, the relation between the matrix field $U(x)$ and the
scalar fields $\pi^a(x)$ takes the form
\bdm U(x) = \exp i \pi(x) \co \hspace{5mm}
\pi(x) = \sum_{a=1}^{3} \pi^a(x) \tau_a \edm
where $\tau_1, \tau_2, \tau_3$ are the Pauli matrices. The
transformation law of these fields may be worked out as follows.

As noted above, the action of $g\in \mbox{G}$ on the element
$n\in \mbox{G/H}$ is
given by $gn=n^\prime h$. In the case under consideration, G consists of
pairs of elements $g=(V_{\indR},V_{\indL})$, while H contains the equal
pairs,
$V_{\indR}=V_{\indL}$. As representative elements of the equivalence classes,
we may choose $n=(U,{\bf 1})$. The transformation law then amounts to
\bdm gn=(V_{\indR},V_{\indL})(U,{\bf 1})=(V_{\indR}U,V_{\indL})=
(V_{\indR}UV_{\indL}^\dagger,{\bf 1})(V_{\indL},V_{\indL})
=n^\prime h \fs\edm
Hence the transformation law of the pion field reads
\bdm\label{sym2} U^\prime(x) = V_{\indR} U(x) V_{\indL}^\dagger\fs\edm
The matrix $U(x)$ thus transforms linearly. Note, however, that the
corresponding transformation law for the pion field $\pi^a(x)$ is
nonlinear --- the matrix $i\pi^a\tau_a$ is the logarithm of the matrix $U(x)$.
As indicated by the above general discussion, the occurrence of
nonlinear realizations of the symmetry group is a characteristic feature of the
effective Lagrangian technique.
%%%%%%%%%%%%%%%%%%%%%%%%%%%%%%%%%%%%%%%%%%%%%%%%%%%%%%%%%%%%%%%%%%%%%%%%
\section{Form of the effective Lagrangian}
\label{form}
Expressed in terms of the field $U(x)$, the effective Lagrangian is of the
form ${\cal L}_{\eff}(U,\partial U,\partial^2U,\ldots)$. Lorentz invariance
implies that
the leading terms in the expansion of this function in
powers of derivatives are of the form
\bdm
{\cal L}_{\eff}=
g_0(U)+g_1(U)\!\times
\raisebox{0.25em}{\fbox{\rule[0.12em]{0mm}{0mm}\,}}\,U
+g_2(U)\!\times\!\partial_\mu U\!\times\!\partial^\mu U
+O(p^4)\fs\edm
The crosses indicate that the coefficients $g_n(U)$ carry indices,
which are to be contracted against those of
$\,\raisebox{0.25em}{\fbox{\rule[0.12em]{0mm}{0mm}\,}}\,U$ and $\partial_\mu
U$. The
remainder contains four or more
derivatives of the pion field.

The first term does not
contain derivatives. The corresponding action is invariant under $U\!
\rightarrow
V_{\indR} UV_{\indL}^\dagger\,$ if and only if $g_0(U)$ is independent of $U$.
Hence the first term is
an irrelevant cosmological constant and may be dropped. In fact, we arrived
at the conclusion that the effective Lagrangian does not contain
interaction terms without derivatives, already in section \ref{lag}; the
above rederivation of this result merely illustrates the efficiency of the
effective Lagrangian technique.

Integrating by parts,
the second term can be transformed into the third one, so that we may drop
$g_1(U)$, too. Without loss of generality we can then write the Lagrangian in
the form $g_2(U) \times \Delta_\mu \times \Delta^\mu$ where
$\Delta_\mu$ stands for $-i U^{-1} \partial_\mu U$. The advantage of the
manipulation is that $\Delta_\mu$ is invariant under $U \rightarrow V_{\indR}
U$, such that only $g_2(U)$ is affected by this operation. The
requirement that the action must remain invariant, therefore,
implies that $g_2(U)$ is independent of U. Finally, under the
transformation $U \rightarrow UV_{\indL}^\dagger$, the traceless quantity
$\Delta_\mu$
transforms according to the representation $D^{(1)}$. Since the product
$D^{(1)}\!\! \times\!
D^{(1)}$ contains the identity only once, there
is a single invariant of order $p^2$,
\be\label{efflag1}
{\cal L}_{\eff}^2 = g\, \mbox{tr} \Delta_\mu \Delta^\mu =
g \,\mbox{tr}\, (\partial_\mu U \partial^\mu U^\dagger)\fs\ee
This shows that the leading term in the derivative expansion of the effective Lagrangian
contains only one free coupling constant, $g$.

Expanding the matrix $U(x) = \exp i \pi(x)$
in powers of the pion field, we obtain
\bdm\label{efflag2}
{\cal L}_{\eff}^2 =
2 g  \partial_\mu \pi^a \partial^\mu \pi^a +
\mbox{$\frac{1}{12}$} g\,\mbox{tr}\,
([\partial_\mu \pi, \pi][ \partial^\mu \pi, \pi]) + \ldots\co
\edm
where interactions involving six or more pion fields are omitted.
The first term represents the kinetic energy of the pion field. To arrive at
the standard normalization of this term, i.e., to insure that the pion
propagator agrees with the one introduced above,
the pion
field must be scaled: the canonical coordinates $\pi^a$ are to be replaced by
$\pi^a/2\sqrt{g}$, such that the expression for the
matrix $U$ takes the form
\bdm
U=\exp\,\left(\frac{i\pi^a\tau_a}{2\sqrt{g}}\right)\fs\edm

The Noether currents associated with the SU(2)$\times$SU(2)
symmetry of the Lagrangian in equation (\ref{efflag1}) are
\bdm
V^\mu_a = ig\, \mbox{tr}\, (\tau_a [\partial^\mu U, U^\dagger])\co\hspace{3em}
A^\mu_a = ig\, \mbox{tr}\, (\tau_a \{ \partial^\mu U, U^\dagger\})\fs\edm
Comparing the expression for the axial current with equation (\ref{eft31}) or
(\ref{effcurrent}), we see that the
coupling constant $g$ is related to the pion decay constant $F$ by $g= F^2/4$.
At leading
order in the derivative expansion, the effective Lagrangian, therefore, only
involves the pion decay constant,
\be\label{efflag}
{\cal L}_{\eff}^2 = \mbox{$\frac{1}{4}$}\,F^2 \,\mbox{tr}\,
(\partial_\mu U \partial^\mu U^\dagger)\co\hspace{3em}
U = \exp \,( i \pi^a \tau_a/F) \fs\ee

The field theory characterized by this Lagrangian is referred to as the nonlinear
$\sigma$-model.
It is well-known that, for $d>2$, this model is not
renormalizable --- taken
by itself, it is not a decent theory.
Actually,
in the above analysis, only the tree graphs of the effective Lagrangian played
a role.
Renormalizability is not an issue which concerns the tree graphs. We will have
occasion to
discuss the significance of loop graphs later on, when we consider the low
energy
expansion beyond leading order. Clearly, the effective Lagrangian must then
also be worked
out beyond the leading term in the derivative expansion. In the
framework of the effective
Lagrangian, the nonlinear $\sigma$-model only represents one building block
of the
construction --- it does not occur by itself. As we will see, the effective
theory as a whole is a perfectly renormalizable scheme.
%%%%%%%%%%%%%%%%%%%%%%%%%%%%%%%%%%%%%%%%%%%%%%%%%%%%%%%%%%%%%%%%%%%%%%%
\section{Geometry and universality}
\label{geo}
The above explicit result for the effective Lagrangian involves the matrix
representation for the pion field. The expression may be rewritten in
terms of the variables $\pi^1,\pi^2,\pi^3$ as follows. The properties of the
Pauli matrices imply
\bdm U={\bf 1}\,\cos\alpha
+i\,\frac{\vec{\tau}\cdot\vec{\pi}}{|\vec{\pi}|}\,
\sin\alpha\co\hspace{4em}
\alpha\equiv\frac{|\vec{\pi}|}{ F} \fs\edm Inserting this in (\ref{efflag}),
the Lagrangian takes the form
\bea
\label{geo1}
{\cal L}_{\eff}^2\al=\al
\mbox{$\frac{1}{2}$}g_{ab}(\pi)\partial_\mu\pi^a\partial^\mu\pi^b\co \no
 g_{ab}(\pi) \al=\al\delta_{ab}\,(\sin\alpha/\alpha)^2+
\frac{\pi^a\pi^b}{\vec{\pi}^{\,2}}\{1-
(\sin\alpha/\alpha)^2\}\fs\nonumber\eea
The expression for $g_{ab}(\pi)$ represents the metric of a sphere of radius
$F$. The effective Lagrangian thus admits the following simple geometric
interpretation.
The pion field variables are the coordinates needed to label the points of
the quotient space G/H = SU(2). Quotient spaces always possess an
intrinsic metric. In the present case, the quotient space even represents a
group. The intrinsic geometry of the group SU(2) is the
one of the three-dimensional unit sphere.
The geometry relevant for the effective Lagrangian coincides with the
intrinsic geometry of the quotient space, except for an
overall normalization factor --- the radius of the relevant sphere is given by
the pion decay constant. As discussed in section \ref{lag}, the expansion of
the metric in powers of the coordinates determines the interaction vertices.
In particular, the four-pion interaction is determined by the second
derivatives
$\partial_{cd}g_{ab}(0)$, i.e., by the curvature of the manifold at the origin.

It is not difficult to understand why the relevant geometry is one of constant
curvature. For the effective Lagrangian to remain invariant under
G = SU(2)$\times$SU(2), the metric occurring therein must admit G as
a group of isometries. In the present case, the transformations of the pion
field represent right- and left-translations on the group SU(2), $U\rightarrow
V_{\indR}UV_{\indL}^\dagger$. For compact groups, this
property fixes the metric uniquely, up to normalization --- the sphere is the
only compact manifold with a six-parameter group of isometries.

The coordinates are a matter of choice.
Replacing the above canonical coordinates by stereographic ones,
the effective Lagrangian simplifies to
\bdm {\cal L}_{\eff}^2=\mbox{$\frac{1}{2}$}\,
\frac{\partial_\mu\vec{\pi}\cdot\partial^\mu\vec{\pi}}
{(1+\frac{1}{4}\,\vec{\pi}^{\,2}/F^2)^{\raisebox{0.2em}{\scriptsize 2}}}=
\mbox{$\frac{1}{2}$}\,\partial_\mu\vec{\pi}\cdot\partial^\mu\vec{\pi}\,
\left(1-\mbox{$\frac{1}{2}$}\vec{\pi}^{\,2}/F^2+\ldots\right)
\edm
I recommend it as an exercise to work out the $\pi\pi$ scattering amplitude
with the above two explicit expressions, to check that, on the mass shell,
the amplitude does not depend on the coordinates used and to verify that
the result agrees with Weinberg's formula \cite{Weinberg 1966}. Off
the mass shell,
the amplitudes, however, differ: for the Green functions of the pion field, the
choice of field variables does matter.
The Green functions of the pion field do not have physical
significance. The effective theory is of interest only as a vehicle, which
allows one to analyze the low energy properties of QCD in an efficient manner.
In QCD, scattering amplitudes and Green functions formed with the
currents
or with the operators $\bar{q}\lambda q,\,\bar{q}\gamma_5\lambda q$ are
meaningful quantities, but a pion field operator only occurs in the
effective theory. For a given choice of the effective field, the
effective
theory does give rise to perfectly unambiguous Green functions also for this
field. In contrast to the results for the scattering amplitudes or for the
Green functions formed with the vector, axial, scalar
and pseudoscalar currents, those of the pion field, however, depend on the
choice of the effective field.

Remarkably, the specific properties of the underlying theory did not
matter in the construction of the effective Lagrangian.
The analysis applies to any theory
for which $\mbox{SU}(2)\times \mbox{SU}(2)$ is spontaneously broken to SU(2).
The explicit expression found for the effective Lagrangian is valid for any
model with these symmetry properties --- the low energy structure is universal.

The extension to three massless flavours is straightforward. In this case, the
Goldstone bosons live in G/H = SU(3). Accordingly, there are eight
pion fields, which may be identified with the canonical coordinates on SU(3),
\bdm\label{uni1}
U = \exp\,(i  \pi^a \lambda_a/F) \co
\edm
where $\lambda_1, \ldots, \lambda_8$ are the Gell-Mann matrices. There is again
only one invariant at order $p^2$,
\be\label{SU(N)}
{\cal L}_{\eff}^2 = \mbox{$\frac{1}{4}$}\,F^2 \,\mbox{tr}\,
(\partial_\mu U \partial^\mu U^\dagger)\fs\ee

In the case of the linear $\sigma$-model with N scalar
fields, the relevant symmetry groups are G= O(N), H = O(N$-$1), such that the
quotient G/H
is the (N$-$1)-dimensional sphere. Accordingly, the pion field is a unit vector
$U^{\ind A}(x)$ with N components. The derivative expansion of the effective
Lagrangian again starts with a term of order $p^2$ and involves a single
coupling constant,
\be\label{SO(N)}
{\cal L}_{\eff}^2 = \mbox{$\frac{1}{2}$}\, F^2 \partial_\mu U^{\ind A}
\partial^\mu U^{\ind A}\fs\ee
The Higgs sector of the Standard Model corresponds to $\mbox{N}=4$.
The corresponding "pion" field
lives in the three-sphere. This manifold may be mapped one-to-one onto the
group SU(2), setting
\bdm
U = {\bf 1} U^0 + i \tau_a U^a\fs\edm
One readily checks that the map takes the Lagrangian in equation (\ref{SU(N)})
into the one in equation (\ref{SO(N)}). This does not represent
a great surprise, because the groups G = O(4) and H = O(3), occurring
in the spontaneous breakdown of the Higgs model, are locally isomorphic to
$\mbox{SU}(2)\times \mbox{SU}(2)$ and to SU(2), respectively. The equivalence
of the two effective theories implies that, at low energies, the Green
functions of the Higgs model and of QCD with two massless flavours are the
same, except for the magnitude of the constant $F$. For QCD, $F \simeq
93$ MeV, while in the case of the Higgs model, $F \simeq 245$ GeV.
%%%%%%%%%%%%%%%%%%%%%%%%%%%%%%%%%%%%%%%%%%%%%%%%%%%%%%%%%%%%%%%%%%%%%%%
\section{Symmetry breaking}
\label{breaking}
In the preceding sections, we have dropped the masses of the $u$- and
$d$-quarks. In their presence, the Lagrangian of the theory is not invariant
under $\mbox{SU(2)}_{\indR}\times\mbox{SU(2)}_{\indL}$, because the mass term
\bdm
{\cal L} = {\cal L}_0 - \bar{q} m q\edm
connects the right- and left-handed components of the quark fields,
\be\label{chirality}
\bar{q} mq = \bar{q}_{\indR} m q_{\indL} + \mbox{h.c.}\fs
\ee
In the notation used here, the quark field $q$ only contains $u$ and $d$ and
$m$ is the matrix
\bdm
m = \pmatrix{m_u &  \cr
 & m_d \cr
}\fs\edm
We may allow for the presence of other quarks, $s,c,\ldots$ They only
appear in ${\cal L}_0$. The only property of ${\cal L}_0$ we will make use
of is that this part of the Lagrangian is
invariant under $\mbox{SU(2)}_{\indR}\times\mbox{SU(2)}_{\indL}$.

It is instructive to compare the QCD Lagrangian with the Hamiltonian of a
Heisenberg ferromagnet,
\bdm
{\bf H} = {\bf H}_0 - \sum_{a} \mu\, \vec{{\bf s}}_a\! \cdot\!
\vec{\,\mbox{H}}\fs \edm
Here, $\vec{{\bf s}}_a$ is the spin associated with lattice site $a$, $\mu$ is
the magnetic moment and $\vec{\,\mbox{H}}$ is an external magnetic field. The
term ${\bf H}_0$ is
invariant under simultaneous O(3) rotations of all spin variables, while the
term which involves the external magnetic field breaks this symmetry. Clearly,
the quark masses $(m_u, m_d)$ play a role analogous to the external magnetic
field and the quark condensate $\lvac \bar{u} u \rvac$, $\lvac\bar{d}
d \rvac$ is analogous to the magnetization. In particular, spontaneous
magnetization at zero external field corresponds to a nonzero value of the
quark condensate in the chiral limit $m_u, m_d \rightarrow 0$.

In the case of the magnet, the symmetry breaking term transforms according to
the spin 1 representation of O(3). The decomposition of the quark mass term
given in equation (\ref{chirality}) shows that this term transforms according
to the representation $D^{(\frac{1}{2}, \frac{1}{2})}$ of
$\mbox{SU}(2)_{\indR}\times
\mbox{SU}(2)_{\indL}$. Equivalently, we may say that the QCD
Lagrangian is invariant under the transformation (\ref{effsy2}) of the
quark fields, provided the mass matrix is transformed accordingly,
\bdm
m \rightarrow V_{\indR} m V_{\indL}^\dagger\fs\edm
In this interpretation, the mass matrix plays the role of a "spurion".

The occurrence of a mass term, of course, modifies the form of the effective
Lagrangian,
\bdm
{\cal L}_{\eff} = {\cal L}_{\eff} (U, \partial U,
\partial ^2 U, \ldots, m)\co\edm
which now remains invariant under the transformation $U(x) \rightarrow
V_{\indR}
U(x)V^\dagger_{\indL}$ of the pion field only if one simultaneously also
transforms
the quark mass matrix in the same manner. The modification of the Lagrangian
generated by the quark masses may be analyzed by expanding in powers of $m$.
The first term in this expansion is the effective Lagrangian of the
massless theory, which we have considered in the preceding sections. The term
linear in $m$ is of the form
\bdm
{\cal L}_{sb} = f(U, \partial U, \ldots) \times m\fs\edm

Next, we observe that derivatives of the pion field are suppressed by powers
of the momenta. At leading order in an expansion in both, powers of $m$
and powers of derivatives, the symmetry breaking term in the effective
Lagrangian reduces to an expression of the form $f(U) \times m$.
Moreover, this expression must be invariant under simultaneous chiral
transformations of the matrices $U$ and $m$. There are only two
independent invariants: $\mbox{tr} (m U^\dagger)$ and its complex
conjugate. Hence the leading symmetry breaking contribution is of the form
\bdm
{\cal L}_{sb} = \mbox{$\frac{1}{2}$}F^2 \{ B \, \mbox{tr} (m
U^\dagger) + B^\star \, \mbox{tr} (m^\dagger U)\}\co \edm
where I have extracted a factor $F^2$ for later convenience. The symmetry
breaking involves a new low energy constant, $B$.
Since only the product $Bm$ matters, the phase of $B$ occurs together
with the phase of the quark mass matrix and is related to the possible
occurrence of a parity violating term of the form $\theta  \, G_{\mu
\nu} \tilde{G}^{\mu \nu}$ in the Lagrangian of QCD. The
fact that
the neutron dipole moment is very small implies that the strong interactions
conserve parity to a very high degree of occuracy. Let us therefore require
that the effective Lagrangian is parity invariant. Using the standard basis,
where the quark mass matrix is diagonal and real, this requirement implies that
$B$ is real (the parity operation sends $\pi$ into $-\pi$ and hence
interchanges $U$ with $U^\dagger)$,
\be\label{breaking1}
{\cal L}_{sb} = \mbox{$\frac{1}{2}$}F^2B\, \mbox{tr}\{ m (U+ U^\dagger)\}\fs
\ee
Since $U$ is an element of SU(2), the sum $U+U^\dagger$ is proportional to the
unit
matrix. Accordingly, the leading contribution to the symmetry breaking part of
the effective Lagrangian only involves the sum $m_u + m_d$ of the two quark
masses and, therefore, conserves isospin --- the breaking of isospin symmetry
generated by the mass difference $m_u - m_d$ only shows up if the low energy
expansion is carried beyond leading order.

Expanding $U = \exp \,( i\, \vec{\pi}\! \cdot\! \vec{\tau}/F)$ in powers of the
pion
field $\vec{\pi}$, the Lagrangian (\ref{breaking1}) gives rise to the following
contributions:
\be\label{breaking2}
{\cal L}_{sb} = (m_u + m_d) B \left\{F^2 - \mbox{$\frac{1}{2}$} \vec{\pi}^{\,2}
+ \mbox{$\frac{1}{24}$}\,\vec{\pi}^{\,4}F^{-2} + \ldots \right\}\fs\ee
Up to a sign, the first term represents the vacuum energy generated by the
symmetry breaking. The second is quadratic in the pion field and, therefore,
amounts to a pion mass term. The remaining contributions show that the symmetry
breaking necessarily also modifies the interaction among the Goldstone bosons.

The derivative of the QCD Hamiltonian with respect to $m_u$ is the operator
$\bar{u}u$. Accordingly, the corresponding derivative of the vacuum energy
represents the vacuum expectation value of $\bar{u} u$. Evaluating this
derivative with the first term in equation (\ref{breaking2}), we obtain
\be\label{breaking3}
\lvac \bar{u} u \rvac = \lvac \bar{d} d \rvac
= - F^2B \{1 + O(m) \}\fs\ee
This shows that the low energy constant $B$ is related to the value
of the
quark condensate. In analyzing the form of the effective Lagrangian, we have
retained only terms linear in the quark masses. The curly bracket
indicates that, in this relation, the higher order terms generate
corrections of $O(m)$.
%%%%%%%%%%%%%%%%%%%%%%%%%%%%%%%%%%%%%%%%%%%%%%%%%%%%%%%%%%%%%%%%%%%%%%%%
\section{Mass of the Goldstone bosons}
\label{mgb}
According to equation (\ref{breaking2}), the pion mass is given by
\be\label{breaking4}
M^2_\pi = (m_u + m_d) B \{ 1 + O(m) \}\fs
\ee
If $m_u$ and $m_d$ are set equal to zero, the pion mass vanishes, as it
should: $\mbox{SU}(2) \times \mbox{SU}(2)$ is then an exact symmetry, such that
the Goldstone
bosons are strictly massless. As long as the symmetry breaking is small, the
Goldstone bosons only pick up a small mass, proportional to the square
root of the symmetry breaking parameter $m_u + m_d$. In accord with the remarks
made above, isospin breaking does not manifest itself at this order of the
expansion --- the masses of $\pi^+, \pi^0$ and $\pi^-$ are the same.

Eliminating the low energy constant $B$, the relations (\ref{breaking3}) and
(\ref{breaking4})
lead to the well-known result of Gell-Mann, Oakes and Renner \cite{GMOR},
\bdm\label{GMORrelation}
F^2 M^2_\pi =  (m_u + m_d) |\lvac \bar{u} u \rvac| + O(m^2)\fs
\edm
The relation shows that the pion mass is determined by the product of the sum
$m_u+m_d$ with the quark condensate. The first factor is a measure of the
{\it explicit} symmetry breaking (which occurs in the Lagrangian of the
theory), while the second is a measure of {\it
spontaneous}
symmetry breaking (for massless quarks, a nonzero value of the
order parameter $\lvac\bar{u}u\rvac$ can only arise if the ground state of the
theory does not possess the same symmetries as the Lagrangian).

The extension from two to three quark flavours is straightforward. The
above analysis
goes through without any essential modifications and leads to an effective
Lagrangian of the same form,
\be\label{mgb1}
{\cal L}_{\eff} = \mbox{$\frac{1}{4}$}F^2 \,\mbox{tr}\, \{ \partial_\mu U
\partial^\mu U^+ + 2 B m (U +U^+) \}\fs\ee
The field $U(x)$ is now an element of SU(3) and describes
eight Goldstone
bosons; $m$ is the diagonal 3$\times$3 matrix formed with
$m_u,m_d$ and $m_s$.

The kinetic
part of the Lagrangian (\ref{mgb1}) is given by the terms quadratic in the
field $\pi^a(x)$, which now carries eight components,
\be\label{mgb2a} {\cal L}_{\eff}^2=\mbox{$\frac{1}{2}$}
\{\partial_\mu\pi^a\partial^\mu\pi^a-B\mbox{tr}(\lambda_a\lambda_bm)
\pi^a\pi^b\}+\ldots\ee
The evaluation of the trace shows that
the masses of those mesons, which carry charge or strangeness, are
given by\footnote{Note that these formulae
concern pure QCD --- the electromagnetic
interaction ge\-ne\-rates corrections of order $e^2$.}
\bea\label{mgb2}
M_{\pi^+}^2\al =\al (m_u + m_d) B + O(m^2)\no
M^2_{K^+}\al =\al (m_u + m_s) B + O(m^2)  \\
M_{K^0}^2\al =\al (m_d + m_s) B + O(m^2) \fs\nonumber\eea
Ignoring the higher order contributions as well as electromagnetic effects,
the above relations may be used to estimate the quark mass ratios from the
observed pion and kaon masses,
\bea\label{mgb3}\frac{m_u}{m_d}\al\simeq\al
\frac{M_{K^+}^2-M_{K^0}^2+M_{\pi^+}^2 }{M_{K^0}^2-M_{K^+}^2+M_{\pi^+}^2}
=0.66\no
\frac{m_s}{m_d}\al\simeq\al
\frac{M_{K^+}^2+M_{K^0}^2-M_{\pi^+}^2 }{M_{K^0}^2-M_{K^+}^2+M_{\pi^+}^2}
=20.1\fs \nonumber
\eea

The mass pattern of the Goldstone bosons breaks SU(3)
symmetry --- the absence of isospin breaking at leading order, observed in the
case of $\mbox{SU}(2)\times \mbox{SU}(2)$, does not repeat itself here. In
fact,
none of the other multiplets shows SU(3) breaking effects comparable to those
seen in the masses of the pseudoscalar octet. At first sight, the fact
that $M_{K}^2$ is 13 times larger than $M_{\pi}^2$ even appears to indicate
that a framework, which assumes
the group G = $\mbox{SU(3)}_{\ind
R}\times\mbox{SU(3)}_{\indR}$
to represent a decent approximate symmetry, is doomed to failure. Note,
however, that in the mass pattern of
the Goldstone bosons, an apparently very strong breaking of SU(3) would occur
even if the quark masses $m_u,m_d,m_s$ where tiny, such that the group G
would represent
an almost perfect symmetry of the QCD Lagrangian: unless the ratios
$m_u:m_d:m_s$ are close to one, the Goldstone bosons pick up very different
masses also in that case.
The point here is that for
SU(3) to be a decent approximate symmetry of the strong interactions, it is
{\it not} necessary that the differences $m_u-m_d$ and $m_d-m_s$ are small
compared to the mean mass $\frac{1}{3}(m_u+m_d+m_s)$. What counts is the
magnitude of the differences
in comparison to the scale of the theory. If $m_u-m_d$ and $m_d-m_s$
are small in this sense, the
symmetry breaking part of the Hamiltonian only generates small corrections. The
quantity to compare $M_K^2-M_\pi^2$ with is not $M_K^2+M_\pi^2$, but the square
of the mass of one of those states, which remain massive when the symmetry
breaking is turned off. A quantitative estimate of the magnitude of
symmetry breaking will be given in section \ref{coupling}.

What surprised us even more, when we realized
that the light quark masses must be very different \cite{GL74/75}, is that
$m_u$ turns out to be quite different from $m_d$, despite the fact that
isospin is an almost perfect symmetry of the strong interactions. The
explanation is the same as for the case of SU(3):
for isospin to be a decent approximate symmetry, it is not necessary that the
difference $m_u-m_d$ is small compared to the sum
$m_u+m_d$. It suffices that the difference is small compared to the scale of
the theory. In the case of the nucleon, e.g., the mass difference
$m_u-m_d$ makes the neutron heavier than the proton by merely 2\permille
(the electromagnetic self-energies are of opposite sign). In the ratio
$(M_{K^0}^2-M_{K^+}^2)/(M_{K^0}^2+M_{K^+}^2)$, the isospin breaking effects are
enhanced, because the denominator is a small quantitiy of order $m$,
but even so, the result is only of the order of 1\%.

For the pions, where one
might have expected the relative magnitude of isospin breaking to be
largest, the matrix elements of the symmetry breaking term turn out to vanish;
there, isospin breaking only shows up at order $(m_u-m_d)^2$. Indeed, we noted
in section \ref{breaking} that the leading term in the derivative
expansion of the effective Lagrangian for two light flavours only involves the
sum $m_u+m_d$ and thus hides the isospin breaking part of the QCD
Hamiltonian. As demonstrated by the mass difference between the $K^0$ and the
$K^+$, this is not the case for three light flavours. The mass term in
equation (\ref{mgb2a}) also induces mixing between the states $\pi^0$ and
$\eta$, through an angle of order $(m_u -m_d)/(m_s-\hat{m})
\ll 1$. The repulsion of the two levels
generates a mass difference between $\pi^0$ and $\pi^+$,
\bdm
M_{\pi^0}^2 \simeq M_{\pi^+}^2 - \mbox{$\frac{1}{4}$}\! \left ( \frac{m_u -
m_d}{m_s - \hat{m}} \right)^2\! (M^2_K - M_\pi^2) \hspace{3em}
\hat{m} \equiv \mbox{$\frac{1}{2}$} (m_u + m_d)\fs\edm
Numerically, this effect is tiny --- the observed mass difference mainly
originates in the electromagnetic self-energy of the charged pion.

Dropping terms of order $(m_u-m_d)^2/(m_s-\hat{m})$, the mass
of the $\eta$ becomes
\bdm M_\eta^2=\mbox{$\frac{2}{3}$}(\hat{m}+2m_s)B+O(m^2)\fs\edm
Accordingly, the squares of the masses obey the
Gell-Mann-Okubo-formula,
\bdm 3M_\eta^2 +M_\pi^2 -2M_{K^+}^2-2M_{K^0}^2 =O\left(m^2,
(m_u-m_d)^2/(m_s-\hat{m})\right)\fs\edm
This relation is satisfied remarkably well, confirming that the group SU(3)
does represesent a decent approximate symmetry.
A quantitative measure for
the magnitude of the symmetry breaking results from a comparison
of two independent determinations of the ratio $m_s/\hat{m}$.
Using the
ratio $(M_{K^0}^2+M_{K^+}^2)/M_{\pi^+}^2$, one obtains $m_s/\hat{m}=24.2$,
while the ratio
$M_\eta^2/M_{\pi^+}^2$ leads to $m_s/\hat{m}=22.7$. If the Gell-Mann-Okubo
formula were exact,
the two results would be the same. The symmetry breaking seen here is
unusually small. In most cases, one finds departures from SU(3) symmetry at the
20-30\% level. A typical example is the ratio of decay constants: the observed
values yield $F_K/F_\pi=1.22$, while exact SU(3) symmetry would require the
two constants to be the same.
%%%%%%%%%%%%%%%%%%%%%%%%%%%%%%%%%%%%%%%%%%%%%%%%%%%%%%%%%%%%%%%%%%%%%%%%%
\section{Currents and external fields}
\label{cur}
In principle, the low energy structure of the current
vertices may be investigated by means of the Noether currents. For the case of
two flavours, the Noether currents belonging to the leading term in the
derivative expansion of the effective Lagrangian
were worked out explicitly in section
\ref{form}. The result specifies the currents as functions of
the pion field and its first derivatives. The formulae given there
represent the leading term in the
derivative expansion of the full effective field theory representation, which
is of the form
\bdm V^\mu_a=V^\mu_a(\pi,\partial\pi,\ldots)\co\hspace{2em}
A^\mu_a=A^\mu_a(\pi,\,\partial\pi,\,\ldots)\fs\edm

For the general
analysis, however, this method is not adequate, because
the low energy expansion of the
Green functions gives rise
to one-particle reducible contributions, which involve more than one current
at the same vertex. Such
vertices
cannot be represented in terms of the above functions,
which describe emission and
absorption of pions by a single current --- they require their own
effective field representation.

The external field method is considerably more efficient, as it treats all of
the vertices
on the same footing. In this approach, one studies the response of the system
to the perturbations generated by suitable external fields. To analyze the
Green functions formed with the vector and axial currents, e.g., one
introduces an external field $v^a_\mu(x)$ coupled to the vector currents as
well as a field $a^a_\mu(x)$ for the axial currents and perturbs the
QCD Lagrangian by a term of the
form
\bdm\label{gf1}   {\cal L}_{\QCD}\rightarrow{\cal L}_{\QCD}+v^a_\mu V^\mu_a
+a^a_\mu A^\mu_a\;\;;\hspace{2em}
V^\mu_a=\bar{q}\gamma^\mu\mbox{$\frac{1}{2}$}\lambda_aq\;,\;\;
A^\mu_a=\bar{q}\gamma^\mu\gamma_5\mbox{$\frac{1}{2}$}\lambda_aq\fs
\edm

The perturbation
generates excitations. Suppose the external fields vanish for $t\rightarrow
\pm\infty$ and assume that, in the remote past, the system was in the
unperturbed
ground state. Denote the vector, which describes this state in the Heisenberg
picture, by $|0\,\mbox{in}\rangle\!\rule[-1mm]{0mm}{5mm}_{\,v,a}\,$.
The effective action is defined as the
logarithm
of the probability amplitude for the system to end up in the ground state for
$t\rightarrow+\infty$, \bdm
\label{eff4}
\exp iS_{\eff}\{v,a\}=\langle 0 \,\mbox{out}\mid 0\,
\mbox{in}\rangle\!\rule[-2mm]{0mm}{5mm}_{\,v,a} \fs
\edm
Perturbation theory shows that this
amplitude is given by the expectation
value (in the unperturbed ground state) of the
time-ordered exponential of the perturbation,
\be \exp iS_{\eff}\{v,a\}=\lvac
T\exp\,i\!\int\!d^4\!x\bar{q}\gamma^\mu
\{v_\mu(x)+\gamma_5a_\mu(x)\}q\rvac\fs\ee
The matrix fields $v_\mu(x),a_\mu(x)$ occurring here are defined by
\bdm v_\mu(x)\equiv \mbox{$\frac{1}{2}$}\lambda_av^a_\mu(x)\hspace{3em}
a_\mu(x)\equiv \mbox{$\frac{1}{2}$}\lambda_aa^a_\mu(x)\fs
 \edm
For definiteness, I consider the case relevant for most of the applications,
where $m_u,m_d$ and $m_s$ are treated as perturbations, retaining the
masses of the remaining, heavy quarks at their physical values. The fields
$v_\mu(x),a_\mu(x)$ then represent hermitean $3\times 3$
matrices. Also, I disregard the singlet currents, taking these matrices
to be traceless.

The above formula shows that the expansion of the effective action in powers of
the external fields yields the Green functions of the currents. The two-point
function of the axial current, e.g., is the coefficient of the term quadratic
in $a_\mu(x)$,
\bdm \exp iS_{\eff}\{v,a\}=1 -\mbox{$\frac{1}{2}$}\!
\int\!d^4\!xd^4\!y \,a^a_\mu(x)a^b_\nu (y)\,
\lvac T A^\mu_a(x)A^\nu_b(y)\rvac\,+\ldots\edm
The advantage of writing the transition amplitude as an exponential is known
from statistical mechanics: the exponent then collects the connected part of
the correlation functions. So, {\it the effective action is the generating
functional of the connected
parts of all of the Green functions formed with the vector and axial currents.}

The presence of external fields in the Lagrangian of the underlying theory, of
course, also shows up in the effective Lagrangian, which now involves
the pion field as well as the external ones,
\bdm {\cal L}_{\eff }={\cal L}_{\eff}(\pi,v,a,\partial\pi,\partial
v,\partial a,\ldots) \fs\edm
The first term in the expansion in powers of $v_\mu,a_\mu$
is the effective Lagrangian considered previously, while the
contributions linear in $v_\mu,a_\mu$ yield the effective field
representations of the currents mentioned above,
\bea {\cal L}_{\eff}(\pi,v,a,\partial\pi,\partial
v,\partial a,\ldots)=\hspace{10em}\no
{\cal L}_{\eff}(\pi,\partial\pi,\ldots)+v_\mu^a
V^\mu_a(\pi,\partial\pi,\ldots)
+a_\mu^aA^\mu_a(\pi,\,\partial\pi,\,\ldots)+O(v^2,va,a^2)\fs
\nonumber\eea
The higher order contributions account for those vertices,
which contain more than one vector or axial current.

The Green functions of the scalar and pseudoscalar currents may also be
generated from the effective action, if we allow the quark mass
matrix to become a space-time dependent field $m(x)$. The QCD
Lagrangian then takes the form
\bdm {\cal L}_{\QCD}={\cal L}_{\QCD}^0 +  \bar{q}\gamma^\mu
\{v_\mu(x)+\gamma_5a_\mu(x)\}q -\bar{q}_{\indR}m(x)q_{\indL}-
\bar{q}_{\indL}m^\dagger(x)q_{\indR}\co\edm
where ${\cal L}_{\QCD}^0$ does not contain external fields and describes the
light quarks as massless particles. The corresponding effective action now also
depends on the external field $m(x)$,
\bdm\label{eff4a}
\exp iS_{\eff}\{v,a,m\}=\langle 0 \,\mbox{out}\mid 0\,
\mbox{in}\rangle\!\rule[-2mm]{0mm}{5mm}_{\,v,a,m} \fs
\edm
The expansion of this functional in powers of $v_\mu(x),a_\mu(x)$ and $m(x)$
yields the Green
functions of the vector, axial, scalar and pseudoscalar currents, {\it in the
limit $m_u\!=\!m_d\!=\!m_s\!=\!0$.} The quark condensate of the massless
theory, e.g.,
is given by the term linear in $m(x)$, all other sources being switched off,
\bdm S_{\eff}\{v,a,m\}=-\!\int\!d^4\!x\,\lvac\bar{q}_{\indR}mq_{\indL}+
\bar{q}_{\indL}m^\dagger q_{\indR}\rvac +\ldots\edm

The same generating functional also contains the Green functions of real QCD.
To extract these, one considers the infinitesimal neighbourhood of the physical
quark mass matrix $m_0$ rather than the vicinity of the point $m=0\,$: Set
$m(x)=m_0+\tilde{m}(x)$ and treat $\tilde{m}(x)$ as a perturbation.
The expansion of the effective action in powers of $v_\mu,a_\mu$ and
$\tilde{m}(x)$ yields the Green functions of the various currents for the case
of physical interest, where the quark masses are different from zero.
%%%%%%%%%%%%%%%%%%%%%%%%%%%%%%%%%%%%%%%%%%%%%%%%%%%%%%%%%%%%%%%%%%%%%%%
\section{Multipion exchange, unitarity and loops}
\label{multi}
Until now, we exclusively dealt with the leading terms of the low energy
expansion. According to the pion pole dominance hypothesis, these are given by
the pole terms due to one-pion exchange. In the language of the effective field
theory, the pole terms arise from the tree graphs.
The hypothesis does not imply,
however, that one-pion exchange dominates to all orders. In fact,
clustering requires that
processes involving the simultaneous exchange of two or more pions between the
same two vertices necessarily also occur. The corresponding contribution
is determined by the
product of the vertices describing emission and absorption,
integrated over the relevant phase space. Since the low energy behaviour of
the vertices is fixed by the leading term in the effective Lagrangian,
the same is true of the multipion exchange
contributions. These processes generate specific low energy
singularities, which
manifest themselves at nonleading orders of the low energy expansion.

Figure \ref{fig2}
indicates some of the effective field theory graphs contributing to
elastic scattering. Note the distinction to the graphs in figure \ref{fig1}:
The vertices occurring there represent full one-particle irreducible
amplitudes, which include contributions
from multipion exchange, while those
in fig.\ref{fig2} correspond to interaction terms
of the effective Lagrangian. In fact, all of the above graphs
depict one-particle irreducible
contributions to the scattering amplitude --- the four-pion vertex marked
with a shaded blob in figs.\ref{fig1}c and \ref{fig1}e includes all of these.

At low
energies, the scattering amplitude is dominated by the tree graph
in fig.\ref{fig2}a. Graphs \ref{fig2}b, \ref{fig2}c and \ref{fig2}d represent
the
exchange of a pair of pions in the $s$-, $t$- and $u$-channel, respectively.
The two pions may undergo a collision under way (\ref{fig2}e) and
contributions involving the exchange of more than two particles also occur
(\ref{fig2}f, \ref{fig2}g). In the language of the effective
field theory, these processes correspond to loop graphs.

Quite generally, graphs involving loops
are essential for the transition amplitudes to conserve probability. Since tree
graphs generate purely real contributions to the T-matrix, they do not
satisfy
the unitarity relation $\mbox{Im}\,T=T^\dagger T$. As pointed out by Lehmann
\cite{Lehmann},
unitarity requires, e.g., that the low energy expansion of the elastic $\pi\pi$
scattering amplitude contains specific contibutions of order $p^4$, which are
not polynomials in the invariants $s$ and $t$, but contain logarithmic branch
points. Within the effective field theory, the branch points
arise from one-loop graphs of the type \ref{fig2}b, \ref{fig2}c and
\ref{fig2}d. Indeed, general\begin{figure}
\begin{center}
\mbox{\epsfxsize=12cm
      \epsfbox{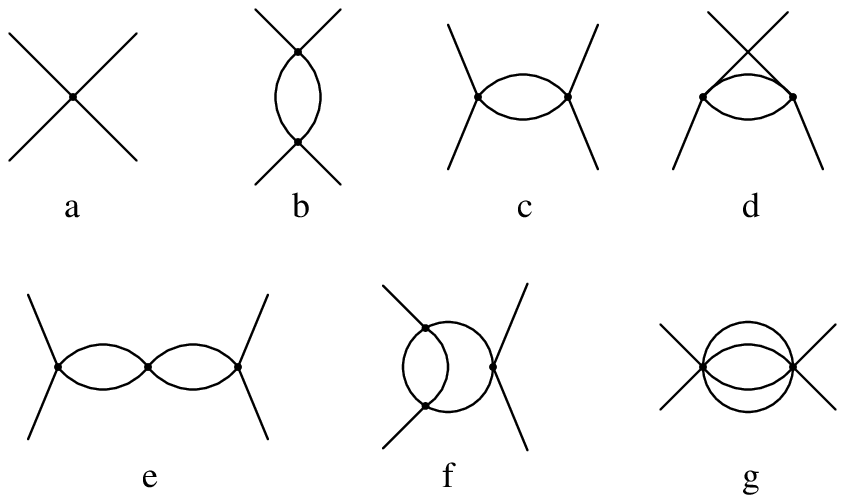} }
\end{center}
\caption{Effective field theory graphs contributing to the
elastic $\pi\pi$ scattering amplitude. The dots represent interaction terms of
the effective Lagrangian.}\label{fig2}\end{figure}\addtocounter{figuren}{1}
kinematics insures that the perturbative expansion of a local field theory
automatically leads to a unitary scattering matrix, provided all of the graphs
are taken into account, including those containing loops. The
corresponding
representation of the effective action is given by the path integral
\bdm
\label{ho5}
e^{i S_{\eff}\{v,a,m\}}= {\cal Z}^{-1}\!\int
[d\pi]\; e^{ i\!\int\!d^4\!x {\cal
L}_{\eff}(\pi,v,a,m,\partial\pi,\partial v,\partial
a,\partial m,\ldots)}\fs \edm The
tree graphs represent the classical limit of this path integral. While
they yield the correct result for the leading terms of the low energy
expansion, the quantum fluctuations described by
graphs containing loops do contribute at nonleading order ---
the pion field of the effective theory is a
quantum field, not a classical one.

Here, the effective Lagrangian method
shows its full strength: the path integral not only yields all of the
pole terms generated by one-pion exchange, but automatically also accounts
for all of the singularities due to the multipion exchange
contributions required by clustering. Because the framework used is
a local field theory, clustering and unitarity are incorporated ab initio.

The above formula is a corner stone of chiral perturbation theory. It
provides the
link between
the underlying and effective theories and is exact, to any finite order
of the low energy expansion. While the
left-hand side represents the generating functional
for the Green functions of the underlying theory, the
right-hand side only involves the effective Lagrangian.

As pointed out by Weinberg \cite{Weinberg1979}, the path integral of the
effective theory
may be evaluated perturbatively, using the momenta, quark masses and external
fields as expansion
parameters.
The resulting perturbation series is identical with the low energy
expansion of the effective action. The higher order terms of the low energy
expansion may be
worked out explicitly by evaluating the path integral to the required accuracy:
the higher orders in the derivative expansion of the effective
Lagrangian need to be
accounted for, as well as graphs involving loops. It is crucial here that, at
low energies, the effective
interaction among the Goldstone bosons is weak (compare section
\ref{strength}). This property insures that the interaction may be accounted
for perturbatively.

In principle, the cuts generated by multipion exchange may also be analyzed
without recourse to an effective Lagrangian, exploiting analyticity and
unitarity and evaluating dispersion relations rather than
loop graphs.
The virtue of the effective
Lagrangian method is that it systematically accounts for all of the
singularities relevant at a given order of the expansion, is straightforward
and free of ambiguities.
%%%%%%%%%%%%%%%%%%%%%%%%%%%%%%%%%%%%%%%%%%%%%%%%%%%%%%%%%%%%%%%%%%%%%
\section{Ward identities}
\label{ward}
One of the virtues of the external field method is that, in this framework, the
Ward identities
take a remarkably simple form. These express
the symmetry properties of the underlying theory in terms of the Green
functions. The Ward identity obeyed by the two-point function formed with
an axial current and a pseudoscalar density, e.g., reads
\bea\label{ward2}
\partial_\mu^x\,\lvac T\;\bar{q}(x)\gamma^\mu\gamma_5
\lambda_aq(x)\;
\bar{q}(y)\gamma_5
\lambda_bq(y)\rvac
=\hspace{8em}\\ i\lvac T\;\bar{q}(x)\gamma_5
\{m,\lambda_a\}q(x)\;\bar{q}(y)\gamma_5
\lambda_bq(y)\rvac -\delta^4(x-y)
\lvac\bar{q}\{\lambda_a,\lambda_b\}q
\rvac\fs\nonumber\eea
The relation may be derived from current conservation and
equal time commutation relations. This derivation, however, leaves to be
desired, because it involves formal manipulations with products of operators
and step functions.
A mathematically satisfactory framework --- which, moreover, yields all
of the Ward identities at once --- is the following.

Consider the Dirac operator, which in the
presence of the external fields introduced above takes the form
\bdm
\label{wi25}
D \;=-\,i\gamma^\mu \{\partial_\mu -i(G_\mu +v_\mu + a_\mu \gamma_5)\}
+ m\mbox{$\frac{1}{2}$}(1-\gamma_5)+
m^\dagger \mbox{$\frac{1}{2}$}(1+\gamma_5) \fs
\edm
The colour field $G_\mu$
is a dynamical field, which mediates the strong
interactions, while the flavour fields $v_\mu,a_\mu$ are classical
auxiliary variables.\footnote{The present discussion concerns QCD; if the
electroweak interactions
are turned on, some of the flavour fields also acquire physical significance.}
In the path integral,
$G_\mu$ is to be integrated over, while $v_\mu,\,a_\mu$ are held fixed.
The colour group acts on the quark and gluon fields,
\bdm
\label{wi27a}
q(x)^{\,\prime}=V_c(x)\,q(x)\;\;,\;\;
G_\mu (x)^{\,\prime} = V_c(x)G_\mu
(x)V_c^{-1}(x)
-i\, \partial_\mu V_c(x) V_c^{-1}(x)
\edm
but leaves the external fields untouched. The flavour group, on the other hand,
acts on the quark fields, through independent rotations of the right- and
left-handed components,
\bdm
\label{wi27}
q_{\indR}(x)^{\,\prime}
=V_{\indR}(x)\,
q_{\indR}(x)\hspace{1em},\hspace{1em}
q_{\indL}(x)^{\,\prime}
=V_{\indL}(x)\,q_{\indL}(x)\co
\edm
leaves the gluons untouched, but affects the external fields. The above
expression for the Dirac operator shows that the linear combinations
\bdm
\label{gu61}
f^{\indR}_\mu =v_\mu +a_\mu\;\;,\;\;f^{\indL}_\mu =v_\mu-a_\mu \edm
transform
like gauge fields of the two factor groups in $\mbox{SU(3)}_{\indR}\!\times\!
\mbox{SU(3)}_{\indL}$, \bea
\label{gu62}
f^{\indR}_\mu (x)^{\,\prime}&\hspace{-0.5em} =&\hspace{-0.5em}
V_{\indR}(x)f^{\indR}_\mu(x) V_{\indR}(x)^{-1}
-i\partial_\mu V_{\indR}(x) V_{\indR}(x)^{-1}\co
\no f^{\indL}_\mu (x)^{\,\prime}&\hspace{-0.5em}
 =&\hspace{-0.5em} V_{\indL}(x)
f^{\indL}_\mu(x) V_{\indL}(x)^{-1}
-i\partial_\mu V_{\indL}(x) V_{\indL}(x)^{-1}\co \eea
while the mass matrix transforms according to
\be
\label{wi27b}
m(x)^{\,\prime}= V_{\indR}(x)m(x)
V_{\indL}^{-1}(x)\fs
\ee

The external fields $v_\mu(x)$ and $a_\mu(x)$ thus promote
the global flavour symmetry of the Lagrangian to a local one, where the
group elements $V_{\indR},V_{\indL}$ may depend on space-time --- as it is the
case with the gauge transformations of colour. This
illustrates the ancient observation of Weyl, according to which any continuous
symmetry may be extended to a local symmetry by introducing suitable gauge
fields: it suffices to replace the ordinary derivatives occurring in the
Lagrangian by covariant ones. In the present case, the relevant covariant
derivatives of the quark fields are
\bdm D_\mu q_{\indR}=(\partial_\mu -iG_\mu
-if_\mu^{\indR})\,q_{\indR}\hspace{3em}
 D_\mu q_{\indL}=(\partial_\mu -iG_\mu -if_\mu^{\indL})\,q_{\indL}\fs
\edm
Since the Lagrangian only involves these covariant derivatives, it is invariant
under the gauge transformations of the fields $q(x),G_\mu(x),f_\mu^{\indR}(x),
f_\mu^{\indL}(x)$ and $m(x)$ specified above.

This line of reasoning is formal, because it deals with the fields as
if they were classical variables. The classical field theory
characterized by a given Lagrangian represents the set of all tree graphs of
the corresponding quantum field theory. The argument just given only shows
that, {\it in the tree graph
approximation}, the vacuum-to-vacuum transition amplitude is gauge invariant.
The full transition amplitude also receives contributions from the
quantum fluctuations of the dynamical variables $G_\mu(x),q(x)$, described by
graphs containing loops. The divergences occurring in these graphs require
the introduction of a cutoff, which modifies the properties of the dynamical
variables and may ruin the symmetries of the Lagrangian. The choice of
the regularization procedure is irrelevant,
in the sense that, for a renormalizable theory, the result is independent
thereof. If there is a regularization, which maintains the
symmetries of the classical theory, then these symmetries also hold at the
level of the quantum theory, but, in
general, this is not the case.

The linear $\sigma$-model
\bdm {\cal L}_{\sigma}=\mbox{$\frac{1}{2}$}\partial_\mu\phi^{\ind A}
\partial^\mu \phi^{\ind A}-\mbox{$\frac{1}{4}$}\lambda\left(\phi^{\ind A}
\phi^{\ind A}-v^2\right)^2\edm
is an example, where
a symmetry preserving regularization
does exist: dimensional regularization. The dynamical variables of
that model are N scalar fields $\phi^{\ind A}(x)$. One may generate the
corresponding Green functions by supplementing the Lagrangian with a term of
the form $m^{\ind A}(x)\phi^{\ind A}(x)$, where $m^{\ind A}(x)$ is an external
field, analogous to the matrix field $m(x)$, needed to generate the Green
functions of the scalar and pseudoscalar currents in QCD. Also,
one may introduce external vector fields coupled to the currents, replacing
the ordinary derivatives with covariant ones,
\bdm D_\mu\phi^{\ind A}=\partial_\mu \phi^{\ind A}-if_\mu^{\ind AB}\phi^{\ind
B}\co\;\;\;f_\mu^{\ind AB}=-f_\mu^{\ind BA}   \fs\edm
The Lagrangian
\bdm {\cal L}_{\sigma}=\mbox{$\frac{1}{2}$}D_\mu\phi^{\ind A}
D^\mu \phi^{\ind A}-\mbox{$\frac{1}{4}$}\lambda\left(\phi^{\ind A}
\phi^{\ind A}-v^2\right)^2 +m^{\ind A}\phi^{\ind A}\edm
is gauge invariant and dimensional regularization maintains this symmetry.
Accordingly, the corresponding vacuum-to-vacuum transition amplitude is gauge
invariant to all orders in the perturbative expansion,
\be\label{ward1} S_{\eff}\{f^\prime,m^\prime\}=S_{\eff}\{f,m\}\fs\ee
The effective action collects all of the Green
functions formed with the currents $\phi^{\ind
A}\partial_\mu\phi^{\ind B}-\phi^{\ind B}\partial_\mu\phi^{\ind A}$
and with the field $\phi^{\ind A}$. The invariance
property (\ref{ward1}) summarizes all of the Ward identities obeyed by the
Green functions. The expansion of this relation in powers of the
external fields shows that, in the linear $\sigma$-model, the formal derivation
of the Ward identities by means of the
equal time commutation relations does lead to the correct result.
%%%%%%%%%%%%%%%%%%%%%%%%%%%%%%%%%%%%%%%%%%%%%%%%%%%%%%%%%%%%%%%%%%%%%%%
\section{Anomalies}
\label{anom}
In fermionic theories with chiral couplings of the gauge fields, where some
of the  vertices involve $\gamma_\mu\gamma_5$ rather than $\gamma_\mu$, the
situation is different. A re\-gu\-larization,
which preserves the symmetries of the Lagrangian with respect to chiral gauge
transformations, does not exist.
In particular, dimensional
regularization fails: $\gamma_5$ cannot be continued in
the dimension. Indeed, the effective action of QCD is not invariant under a
gauge transformation of the external fields, \bdm
S_{\eff}\{v^\prime,a^\prime,m^\prime\}\neq
S_{\eff}\{v,a,m\}\co\edm
because the Ward identities pick up extra contributions,
generated
by loop graphs. The formal derivation of the Ward identities, based on the
equal time
commutation relations, misses these. The extra terms are referred to as {\it
anomalies}. The problem does not concern the interaction of the quarks with the
dynamical field
$G_\mu$, which is vector-like: the quantum fluctuations preserve
the symmetry with respect to {\it colour}
gauge transformations. The problem is caused by the perturbations
generated by the external field $a_\mu$,
whose interaction with the quarks distinguishes the right- and left-handed
components. The effective action fails to be invariant under the gauge group of
{\it flavour}.

We encountered the phenomenon already
in section \ref{qcdmq},
where we noted that one of the global U(1)-symmetries of the massless theory is
ruined by an anomaly: The Ward identities obeyed by
the singlet axial current contain an anomalous contribution proportional to
$\epsilon^{\mu \nu\rho\sigma}G_{\mu\nu}G_{\rho\sigma}$. In the presence of
external fields, analogous terms built with $v_\mu$
and $a_\mu$ also occur, such as
$\epsilon^{\mu \nu\rho\sigma}\partial_\mu v_\nu\partial_\rho v_\sigma$ or
$\epsilon^{\mu \nu\rho\sigma}\partial_\mu a_\nu\partial_\rho a_\sigma$.
The problem arises from
fermionic one-loop graphs,
involving three, four or five vertices, at which the quark
emits one of the fields
$G_\mu, v_\mu$ or $a_\mu$.

The external fields $v_\mu,a_\mu$ entering the generating functional
$S_{\eff}\{v,a,m\}$ introduced above are traceless; the
effective action only collects
the Green functions formed with the currents of
SU(3)$_{\indR}\times$SU(3)$_{\indL}$ and does not
contain those involving the singlet currents. The anomaly occurring in
the Ward identities for that current does, therefore, not concern us here.
It is not difficult to see that, if the flavour fields are traceless,
triangle graphs
involving gluons as well as $v_\mu$ or $a_\mu$ vanish upon performing the
traces over colour and flavour. More generally, anomalous contributions
involving gluons do then not arise. This is the basis of the
nonrenormalization theorem of Adler and Bardeen \cite{Adler Bardeen}, which
states that the change in the effective action,
produced by a chiral rotation of the flavour fields,
can be given explicitly, to all orders. Under an
infinitesimal chiral rotation, \bdm
\label{wi31}
V_{\indR}(x) =1 + i\alpha(x)+i\beta(x)\co\;\;\;\;
V_{\indL}(x) = 1+i\alpha(x)-i\beta(x)\co
\edm
the external fields undergo the gauge transformation
\bea
\label{wi35}
\delta v_\mu &=& \partial_\mu \alpha + i[\alpha,v_\mu ] +i[\beta,a_\mu ]
\co\;\;\;
\delta a_\mu = \partial_\mu \beta + i[\alpha,a_\mu ] +i[\beta,v_\mu ]\co \no
\delta m&=& i(\alpha +\beta)m-im(\alpha -\beta)
\fs \nonumber
\eea
The corresponding change in the effective action involves the difference
$\beta$ between $V_{\indR}$ and $V_{\indL}$,
\be
\label{wi36}
 S_{\eff}\{v^\prime,a^\prime,m^\prime  \} = S_{\eff}\{v,a,m  \} -\int\!\!
d^4\!x\, \mbox{tr}\{\beta(x)\Omega(x)\}+O(\beta^2)\fs
\ee
The explicit expression for $\Omega$ is proportional to the number
of colours and exclusively contains the external vector and axial vector
fields, \be
\label{wi34}
\Omega = \frac{\mbox{N}_c}{4\pi^2}\varepsilon^{\mu \nu \rho
\sigma}\{\partial_\mu
v_\nu \partial_\rho v_\sigma +\mbox{$\frac{1}{3}$}\partial_\mu a_\nu
\partial_\rho a_\sigma +\ldots \}\fs
\ee
The terms listed are those responsible for the Adler-Bell-Jackiw
anomaly of the triangle graphs.
In addition, $\Omega$ also contains terms
involving three or four external fields, describing
the anomalies in quark loops with 4 or 5 external field vertices.
The specific form of these is not relevant here --- the main point is that the
change in the effective action is an explicitly known expression, which is of
geometrical nature and does not depend on the gluon field $G_\mu$, nor on the
QCD coupling constant, nor on the masses of the heavy quarks, nor on
the external field $m(x)$, which contains the masses of the light quarks.

The transformation law (\ref{wi36}) represents a compact summary of all of the
Ward identities obeyed by the Green functions of QCD. It states, in
particular, that anomalies only occur in the 3-, 4- and 5-point
functions, formed exclusively with the currents. Moreover, since the quantity
$\Omega$ is proportional to $\epsilon^{\mu\nu\rho\sigma}$, the Green
function in question must contain an odd number of
axial currents. The three-point function $\lvac T A_\mu V_\rho V_\sigma\rvac$
is the most prominent example; the anomalous contribution
in the Ward identities obeyed by this quantity plays a central role in the
decay $\pi^0\rightarrow\gamma\gamma$. The two-point function
formed with an axial current and a pseudoscalar density, on the other
hand,
obeys an anomaly free Ward identity: the one written
down in equation (\ref{ward2}).
%%%%%%%%%%%%%%%%%%%%%%%%%%%%%%%%%%%%%%%%%%%%%%%%%%%%%%%%%%%%%%%%%%%%%%%%%
\section{Higher orders in ${\cal L}_{\eff}$}
\label{high}
The effective Lagrangian collects all of the vertices, purely pion
interactions, as well as those describing the interaction with the
external fields $v_\mu(x),a_\mu(x),m(x)$. To order the various
vertices, we first observe that the quark masses
generate a pion mass proportional to the square root of $m_u+m_d$. When
analyzing on-shell matrix elements such as scattering amplitudes, the momenta
obey the condition $p^2=M_\pi^2\propto (m_u+m_d)$. A coherent bookkeeping of
the powers of momenta thus requires that the quark masses are counted as
perturbations of order $p^2$. We will use this counting of powers also for
the corresponding external field, treating the matrix $m(x)$ as a
quantity of $O(p^2)$
\bdm m(x)\sim p^2\fs\edm
Concerning $v_\mu(x)$ and $a_\mu(x)$, we note
that gauge transformations modify the combinations $v_\mu\pm a_\mu$ by a
term
involving the first derivative of the matrices $V_{\indR}(x),V_{\indL}(x)$.
It is therefore convenient to count these fields as quantities
of the same order as the derivative,
\bdm v_\mu(x),a_\mu(x)\sim p\fs\edm
This bookkeeping insures that the Ward identities
relate terms of the same order in the low energy expansion.

Lorentz invariance then implies
that, as before, the expansion of the effective Lagrangian in the number of
derivatives and external fields only involves even powers,
\be\label{high1} {\cal L}_{\eff}={\cal L}_{\eff}^2+{\cal L}_{\eff}^4+{\cal
L}_{\eff}^6+\ldots\ee
If the fields $v_\mu,a_\mu$ are turned off, the leading term in this expansion
is given by the two pieces worked out in the preceding sections,
\bdm
{\cal L}_{\eff}^2 = \mbox{$\frac{1}{4}$}\,F^2\,\mbox{tr}\,
(\partial_\mu U \partial^\mu U^\dagger)+
\mbox{$\frac{1}{2}$}F^2B\,\mbox{tr}(mU^\dagger+Um^\dagger)\fs\edm
The first one is of order $p^2$, because it involves two derivatives of the
pion field. The contribution generated by the symmetry breaking
counts as a term of the same order, because it is
linear in $m=O(p^2)$. The next term in the expansion, ${\cal L}^4_{\eff}$,
contains
contributions with four derivatives, terms with two derivatives and one factor
of $m$, as well as contributions which are quadratic in $m$, etc.

If the fields $v_\mu,a_\mu$ are turned on, each one of the
terms ${\cal L}^2_{\eff},{\cal L}^4_{\eff},\ldots\,$ picks up additional
contributions. Their form is very strongly constrained by the Ward identities.
Suppose for a moment that we are dealing with the effective Lagrangian of an
anomaly free theory, such as the linear $\sigma$-model. As discussed above,
the Ward identitites are then equivalent to the statement that
the generating functional $S_{\eff}\{v,a,m\}$ is invariant under a gauge
transformation of the fields $v_\mu$, $a_\mu,m$. The implications for
the
form of the effective Lagrangian are remarkably simple: The invariance theorem
asserts that $S_{\eff}\{v,a,m\}$ is gauge invariant if and only if
the effective Lagrangian is (for a proof, I again refer to
\cite{found}).

The invariance theorem also covers the
case of theories like QCD, where the Ward identities contain anomalous
contributions. In that case, the effective Lagrangian consists of two parts,
\bdm {\cal L}_{\eff}=
\stackrel{\,\rule{0.5em}{0.03em}}{\cal L}_{\eff} +\,\LWZ\fs\edm
The quantity $\LWZ$ is the famous Wess-Zumino term,
an explicitly known
expression, involving the pion field $U(x)$ as well as the external
fields $v_\mu(x)$ and $a_\mu(x)$. In the bookkeeping introduced above, the
Wess-Zumino Term is a term of order $p^4$.
The theorem asserts that, once this contribution is
removed, the remainder, $\stackrel{\,\rule{0.5em}{0.03em}}{\cal L}_{\eff}$, is
gauge
invariant, i.e., has the same properties as the effective Lagrangian of an
anomaly free theory. This implies, that, in the derivative expansion
(\ref{high1}), all of the terms except ${\cal L}_{\eff}^4$ are gauge invariant.

If the Wess-Zumino term is dropped,
the path integral of the effective theory yields a gauge invariant effective
action, $\delta S_{\eff}=0$. By construction, the term $\LWZ$
modifies the
effective action in such a manner that it instead obeys the transformation law
$\delta S_{\eff}=-\!\int\!d^4\!x\mbox{tr}\{\beta\Omega\}$. Accordingly, the
Green functions generated by this functional automatically
obey modified Ward identities, which contain the relevant
anomalous contributions.

It is not difficult to convert the
above explicit expression for ${\cal L}^2_{\eff}$ into a gauge invariant one:
it suffices to replace the ordinary derivatives of the pion field by covariant
derivatives. The form of the covariant derivative immediately follows from the
transformation law for the pion field, derived in section \ref{tp},
\bdm U^\prime(x)=V_{\indR}(x)U(x)V_{\indL}^\dagger(x)\fs\edm
The transformation law (\ref{gu62}) of the external fields shows that the
quantity
\be D_\mu U =\partial_\mu U-if_\mu^{\indR}U+i\,U\!f_\mu^{\indL}\ee
transforms in the same manner as the field $U$. Hence the expression
\be
{\cal L}_{\eff}^2 = \mbox{$\frac{1}{4}$}\,F^2\,\mbox{tr}\,
(D_\mu U D^\mu U^\dagger)+
\mbox{$\frac{1}{2}$}F^2B\,\mbox{tr}(mU^\dagger+Um^\dagger)\}\ee
is invariant under a gauge transformation of the fields. Indeed, it represents
the most general expression of order $p^2$ with this property.

At order $p^4$, there are quite a few independent invariants, e.g.
\bea{\cal L}_{\eff}^4\al=\al L_1\langle D_\mu U D^\mu
U^\dagger\rangle^2
+ L_5\langle D_\mu U D^\mu U^\dagger (\chi U^\dagger +U\chi^\dagger)\rangle
+L_7\langle\chi U^\dagger-U\chi^\dagger\rangle^2\no
\al-\al iL_9\langle f_{\mu\nu}^{\indR}D_\mu U D^\nu U^\dagger
+f_{\mu\nu}^{\indL}D_\mu U^\dagger D^\nu U \rangle +\ldots
+\,\LWZ\co \nonumber\eea
where the external field $m(x)$ has been replaced by $\chi(x)\equiv 2Bm(x)$.
The symbol $\langle A\rangle$ denotes the trace of the $3\times 3$ matrix
$A$ and $f_{\mu\nu}^{\indR},f_{\mu\nu}^{\indL}$ stand for the field
strengths belonging to $f_\mu^{\indR},f_\mu^{\indL}$, respectively. All of
the above terms (for a full list, see \cite{GLNP}) are manifestly gauge
invariant, except for $\LWZ$. The effective coupling
constants $L_1,L_2,\ldots$ are the analogues of the two quantities
$F,B$, which specify the effective Lagrangian at leading order of the
derivative expansion.
%%%%%%%%%%%%%%%%%%%%%%%%%%%%%%%%%%%%%%%%%%%%%%%%%%%%%%%%%%%%%%%%%%%%%%%
\section{Renormalizability}
\label{ren}
As mentioned in section \ref{multi}, the low energy expansion of the
path integral over the pion field may be worked out by means
of perturbation theory. The leading term of the expansion is given by the
tree graphs. In the case of the
$\pi\pi$ scattering amplitude, e.g., the tree graph contribution, shown in
fig.\ref{fig2}a, is of order $p^2$. The one loop graphs of
fig.\ref{fig2}b,\ref{fig2}c and \ref{fig2}d generate a contribution of order
$p^4$, while graphs with two loops only contribute at order $p^6$.
More generally, graphs containing a different number of loops occur at
different orders of
the low energy expansion: in $d$ dimensions, graphs with $\ell$ loops are
suppressed
compared to the tree graphs by the power $[p^{d-2}]^\ell$. The rule is readily
checked for
individual graphs such as those shown in fig.\ref{fig2}. The loop integrals
are homogeneous
functions of the external momenta and of the pion mass, which enters through
the propagators. The
degree of homogeneity is determined by the dimension of the integral, which in
turn is
fixed by the overall power of the pion decay constant, arising from the various
vertices.
A more thorough discussion of the issue can be found in ref.\cite{Weinberg1979}.

As discussed above, the Lagrangian ${\cal L}^2_{\eff}$ is not the full story
--- graphs
involving vertices of ${\cal L}^4_{\eff}, {\cal L}^6_{\eff}, \ldots$ also need
to be taken into
account. In the case of the $\pi \pi$ scattering amplitude, graphs
containing $\ell$ loops
are of order $p^{2+\ell(d-2)}$, provided they exclusively involve vertices of
${\cal L}^2_{\eff}$. Graphs containing one vertex of ${\cal L}^4_{\eff} \,
({\cal L}^6_{\eff})$ are smaller
by one (two) powers of $p^2$. Hence, to evaluate the scattering amplitude in
four
dimensions to order $p^4$, we need to work out the tree and one-loop graphs of
${\cal
L}^2_{\eff}$ and add the tree graphs with
one vertex
from ${\cal L}^4_{\eff}$. Higher orders in the derivative expansion of the
effective
Lagrangian and two-loop graphs only start contributing at order $p^6$.

Note that the graphs can be ordered by counting powers of the momentum only if
$d>2$. In
two dimensions, the constant $F$ is dimensionless and the degree of
homogeneity is therefore
independent of the number of loops. In $d=2$, the Lagrangian ${\cal
L}^2_{\eff}$ taken by
itself specifies a decent, renormalizable theory, which moreover is
asymptotically free and thus shares the qualitative properties of
four-dimensional nonabelian gauge theories. In particular, the low energy
structure of the theory cannot be analyzed perturbatively. (Incidentally,
supplementing
${\cal L}^2_{\eff}$ by the Wess-Zumino term, one arrives at a two-dimensional
field theory with
very peculiar properties: the Wess-Zumino-Novikov-Witten model. In this model,
the
coupling constant $F$ can be tuned in such a fashion that the $\beta$-function
vanishes - the theory becomes conformally invariant.)

In $d=4$, the Lagrangian ${\cal L}^2_{\eff}$ by itself is meaningless, but
taken together with
the infinite string of higher order terms ${\cal L}^4_{\eff}, {\cal
L}^6_{\eff}, \ldots\;,$ it does
specify a renormalizable framework. If one disregards from those
vertices, which involve the
tensor $\epsilon^{\mu\nu\rho\sigma}$, one may regularize the
loop integrals
by means of dimensional regularization, which preserves the
symmetries of the Lagrangian. The poles occurring at $d=4$ then only
require counter terms, which
are Lorentz invariant and symmetric under $\mbox{SU(3)}_{\indR}
\times\mbox{SU(3)}_{\indL}$. By construction, the full effective Lagrangian
contains all terms permitted by this symmetry. The divergences may, therefore,
be absorbed in
a renormalization of the coupling constants occurring in the Lagrangian. In
particular,
the divergences contained in the one-loop graphs are absorbed in a renormalization of the
coupling constants $L_1,L_2, \ldots\;,$ occurring in ${\cal L}^4_{\eff}$.
Dimensional regularization also takes care of a
technical complication, connected with
the fact that the effective Lagrangian contains derivative couplings. This
property implies that the measure occurring in the functional integral does not
coincide with the standard translation invariant measure on the space of the
pion fields. In general, the measure generates additional contributions
involving power divergences, such as $\delta(0) \sim \Lambda^4$. In dimensional
regularization, however, power divergences do not occur (in particular,
$\delta (0)$ vanishes)
and the complications associated with the measure can simply be
ignored.\footnote{
For a more detailed discussion and references to the literature, see, e.g.
\cite{GLNP}.}
%%%%%%%%%%%%%%%%%%%%%%%%%%%%%%%%%%%%%%%%%%%%%%%%%%%%%%%%%%%%%%%%%%%%%%%%%
\section{Illustration: electromagnetic form factor}
\label{ff}
As an illustration of the method, let us return to the e.m. form factor of
the pion, considered in section
\ref{intro}. With the machinery developed above, it is straightforward to
calculate this quantity to first nonleading order in the low energy
expansion. To keep these notes within bounds, I do not describe the calculation
here, but refer to the literature \cite{GLNP}. Other sample calculations may be
found in the reviews quoted in the introduction. The result is of the form
\be\label{ff1} f_{\pi^+} (t)  =
1 + \frac{t}{F^2} \{2L_9 + 2 \phi(t,M_\pi) + \phi(t,M_K)\}
+ O(t^2, t m)\ee
In this case, the leading term of the expansion is trivial, as it represents
the charge of the particle, $f_{\pi^+}(0)=1$. At first nonleading order, there
are two types of contributions: (i) The term proportional to $L_9$ evidently
comes from a tree graph containing one vertex from ${\cal L}_{\eff}^4$; it is
linear
in the momentum transfer $t$. (ii) The functions $\phi(t,M_\pi)$ and
$\phi(t,M_K)$ are generated by one-loop graphs, which exclusively
involve vertices from ${\cal L}_{\eff}^2$; they are nontrivial functions of
$t$, containing branch cuts, starting at
$t=4M_\pi^2$ and $t=4M_K^2$. In dispersive language, the cuts are generated
by $\pi\pi$ and
$K\bar{K}$ intermediate states.

The loop integrals diverge. In dimensional regularization, the function
$\phi(t,M)$ contains a pole at $d=4$. The residue of the pole is momentum
independent --- the quantity $\phi(t,M)-\phi(0,M)$ tends to a finite limit
when $d\rightarrow 4$. Accordingly, the divergence may be absorbed in a
suitable renormalization of the coupling constant $L_9$. The result for the
form factor is a finite expression, which is independent of the
regularization used, but involves an effective coupling constant.

The result shows that chiral symmetry does not determine the pion charge
radius: its
magnitude depends on the value of the coupling constant $L_9$ --- the effective
Lagrangian is consistent with chiral symmetry for any value of the coupling
constants.
The symmetry,
however, {\it relates} different observables. The slope of the $K_{l_3}$ form
factor $f_+(t)$, e.g., is also fixed by $L_9$. The experimental value of this
slope, $\lambda_+ = 0.030$, can therefore be used to first determine the
magnitude of $L_9$ and then to calculate the pion charge radius. This gives
$\langle r^2\rangle_{\pi^+} = 0.42$ fm$^2$, to be compared with the
experimental result, 0.44 fm$^2$.

In the case of the neutral kaon, the analogous representation
reads \be
f_{K^0} (t) = \frac{t}{F^2} \{ - \phi_\pi (t) + \phi_K (t) \} + O(t^2, t
m).
\ee
A term of order one does not occur here, because the charge vanishes and there
is no contribution from ${\cal L}^4_{\eff}$, either. Chiral
perturbation theory thus provides a parameter free prediction in terms of the
one loop integrals $\phi_\pi(t), \phi_K(t)$. In particular, the slope
of the form factor is given by \cite{GLNP}
\be
\langle r^2\rangle_{K^0} = - \frac{1}{16\pi^2F^2} \ln \frac{M_K}{M_\pi} = -
0.04\; \mbox{fm}^2\co
\ee
to be compared with the experimental value $- 0.054 \pm 0.026$ fm$^2$.
This result represents the first prediction of this type
--- in the meantime, similar parameter free one-loop predictions have been
discovered for
quite a few other observables.
%%%%%%%%%%%%%%%%%%%%%%%%%%%%%%%%%%%%%%%%%%%%%%%%%%%%%%%%%%%%%%%%%%%
\section{Magnitude of the coupling constants}
\label{coupling}
One of the main problems encountered in the effective Lagrangian
approach is the occurrence of an entire fauna of effective coupling constants.
If these constants are treated as totally arbitrary parameters, the predictive
power of the method is equal to zero --- as a bare minimum, an estimate of
their order of magnitude is needed.

In principle, the effective coupling
constants $F, B, L_1, L_2, \ldots$ are calculable. They do not depend on the
light quark masses, but are determined by the scale $\Lambda_{\QCD}$ and by the
masses of the heavy quarks. The available,
admittedly crude evaluations of $F$ and $B$
on the lattice demonstrate that the calculation is even feasible in practice.
As
discussed above, the coupling constants $L_1, L_2, \ldots$ are renormalized by
the logarithmic divergences occurring in the one loop graphs. This property
sheds considerable light on the structure of the chiral expansion and provides
a rough estimate for the order of magnitude of the effective coupling
constants \cite{Georgi Soldate}. The point is that the contributions generated
by the loop
graphs are smaller than the leading (tree graph) contribution only for momenta
in the range $\mid p \mid \raisebox{-0.3em}{$\stackrel{<}{\sim}$}\,
\Lambda_\chi$,
where  \be\label{chiral scale}
\Lambda_\chi \equiv 4 \pi F/ \sqrt{N_f}
\end{equation}
is the scale occurring in the coefficient of the logarithmic divergence ($N_f$
is the number of light quark flavours). This indicates that the derivative
expansion is an expansion in powers of $(p/\Lambda_\chi)^2$, with coefficients
of order one. The stability argument also applies to the expansion in powers of
$m_u, m_d$ and $m_s$, indicating that the relevant expansion parameter is given
by $(M_\pi/ \Lambda_\chi)^2$ and $(M_K/\Lambda_\chi)^2$, respectively.

A more quantitative picture may be obtained along the following lines. Consider
again the e.m. form factor of the pion and compare the chiral representation
(\ref{ff1}) with the dispersion relation
\bdm
f_{\pi^+}(t) = \frac{1}{\pi} \int^{\infty}_{4M^2_\pi} \frac{dt'}{t'-t}
\mbox{Im} f_{\pi^+} (t')\fs
\edm
In this relation, the contributions $\phi_\pi, \phi_K$ from the one loop graphs
of $\chi$PT correspond to $\pi \pi$ and $K \bar{K}$ intermediate states. To
leading order in the chiral expansion, the corresponding imaginary parts are
slowly rising functions of $t$. The most prominent contribution on the rhs,
however, stems from the region of the $\rho$-resonance which nearly saturates
the integral: the vector meson dominance formula, $f_{\pi^+} (t) =
(1-t/M_\rho^2)^{-1}$, which results if all other contributions are dropped,
provides a perfectly decent representation of the form factor for small values
of $t$. In particular, this formula predicts $\langle r^2\rangle_{\pi^+} =
0.39$ fm$^2$,
in satisfactory agreement with observation (0.44 fm$^2$). This implies that the
effective coupling constant $L_9$ is approximately given by \cite{GLNP}
\be
L_9 = \frac{F^2}{2M^2_\rho}\fs
\end{equation}
In the channel under consideration, the pole due to $\rho$ exchange thus
represents the dominating low energy singularity --- the $\pi \pi$ and $K
\bar{K}$ cuts merely generate a small correction. More generally, the validity
of the vector meson dominance formula shows that, for the e.m. form factor, the
scale of the derivative expansion is set by $M_\rho = 770$ MeV.

Analogous estimates may be given for all effective coupling constants at order
$p^4$, saturating suitable dispersion relations with contributions from
resonances \cite{Ecker,Florida}, e.g.
\bdm\label{L57}
L_5 = \frac{F^2}{4M^2_S}\co\hspace{3em}
L_7 = - \frac{F^2}{48M^2_{\eta'}}\co\edm
where $M_S \simeq 980$ MeV and $M_{\eta'} = 958$ MeV are the masses of the
scalar octet and pseudoscalar singlet, respectively. In all those cases, where
direct phenomenological information is available, these estimates do
remarkably well.
I conclude that the observed low energy structure is dominated
by the poles and cuts generated by the lightest particles --- hardly a
surprise.

The effective theory is constructed on the asymptotic states of QCD. In the
sector with zero baryon number, charm, beauty, $\ldots\,,$ the
Goldstone bosons form a complete set of such states, all other mesons
being unstable against decay into these (strictly speaking, the
$\eta$ occurs among the asymptotic states only for
$m_d=m_u$; it must be included among the degrees of freedom of the
effective theory, nevertheless, because the masses of the light quarks are
treated as a perturbation --- in massless QCD, the poles
generated by the exchange of this particle occur at $p=0$).
The Goldstone degrees of freedom are explicitly accounted for
in the effective theory --- they represent the dynamical variables. All other
levels manifest themselves only indirectly, through the values of the effective
coupling constants. In particular, low lying levels such as the $\rho$
generate relatively small energy denominators, giving rise to relatively large
contributions to some of these
coupling constants.

In some channels, the scale of the chiral expansion is set by $M_\rho$, in
others by the masses of the scalar or pseudoscalar resonances occurring around
1 GeV. This confirms the rough estimate (\ref{chiral scale}). The cuts
generated
by Goldstone pairs are significant in some cases and are negligible in others,
depending on the numerical value of the relevant Clebsch-Gordan coefficient. If
this coefficient turns out to be large, the coupling constant in question is
sensitive to the renormalization scale used in the loop graphs. The
corresponding pole dominance formula is then somewhat fuzzy, because the
prediction depends on how the resonance is split from the continuum underneath
it.

The quantitative estimates of the effective couplings given above
explain why
it is justified to treat $m_s$ as a perturbation. At order $p^4$, the symmetry
breaking part of the effective Lagrangian is determined by the
constants $L_4, \ldots, L_8$. These constants are immune to the low energy
singularities generated by spin 1 resonances, but are affected by the exchange
of scalar or pseudoscalar particles. Their magnitude is, therefore, determined
by the scale $M_S \simeq M_{\eta'} \simeq 1$ GeV. Accordingly,
the expansion in powers of $m_s$ is controlled by the parameter $(M_K/M_S)^2
\simeq \frac{1}{4}$. The asymmetry in the decay constants, e.g., is
determined by $L_5$. The estimate of this coupling constant given above yields
\bdm\label{fk/fpi}
\frac{F_K}{F_\pi} = 1+\frac{M^2_K - M^2_\pi}{M^2_S} + \chi\mbox{logs}
+O(m^2)\co \edm
where the term $"\chi\mbox{logs}"$ stands for the chiral logarithms generated
by the one loop graphs.
This shows that the breaking of the chiral and eightfold way symmetries is
controlled by the mass ratio of the Goldstone bosons to the non-Goldstone
states of spin zero. In $\chi$PT, the observation that the Goldstones are
the lightest hadrons thus acquires quantitative significance: For momentum
independent quantities such as masses, decay constants, charge radii
or scattering lengths, the magnitude of consecutive orders
in the chiral perturbation series is determined by the square of the above mass
ratio.

With this remark, I close the present lecture notes, which concern the
foundations of the method. Plenty of applications
are described in the literature and several different directions of research
are currently under active investigation --- the references
quoted in the introduction provide a rough orientation.\\

It is a pleasure to thank Victoria Herscovitz, C\'{e}sar Vasconcellos,
Jos\'{e} de S\'{a} Borges, Erasmo Ferreira and Juan Mignaco for their warm
hospitality during a most enjoyable stay in Brasil.
%%%%%%%%%%%%%%%%%%%%%%%%%%%%%%%%%%%%%%%%%%%%%%%%%%%%%%%%%%%%%%%%%%%%%%%

\end{document}